\documentclass[11pt]{article}
\pdfoutput=1

\usepackage{amsmath,amsfonts,enumerate,array}
\usepackage{graphicx}
\usepackage[usenames]{color}
\usepackage[english]{babel}
\usepackage{fancybox}
\usepackage{chngpage} 

\newcommand{\captionfonts}{\small}

\makeatletter  
\long\def\@makecaption#1#2{%
  \vskip\abovecaptionskip
  \sbox\@tempboxa{{\captionfonts #1: #2}}%
 \ifdim \wd\@tempboxa >\hsize
    {\captionfonts #1: #2\par}
  \else
    \hbox to\hsize{\hfil\box\@tempboxa\hfil}%
  \fi
  \vskip\belowcaptionskip}
\makeatother   

\usepackage{mciteplus}

\usepackage[colorlinks=true,      linkcolor=blue,      urlcolor=blue,      filecolor=blue,      citecolor=blue,
      pdfstartview=FitH,     pdfpagemode=UseNone,      bookmarksopen=true]{hyperref}
\usepackage[all]{hypcap}     

\begin{document}

\numberwithin{equation}{section}


\mathchardef\mhyphen="2D


\newcommand{\be}{\begin{equation}}
\newcommand{\ee}{\end{equation}}
\newcommand{\bea}{\begin{eqnarray}\displaystyle}
\newcommand{\eea}{\end{eqnarray}}
\newcommand{\nnm}{\nonumber}
\newcommand{\nn}{\nonumber}

\def\eq#1{(\ref{#1})}
\newcommand{\secn}[1]{Section~\ref{#1}}

\newcommand{\tbl}[1]{Table~\ref{#1}}
\newcommand{\fig}{Fig.~\ref}

\def\beq{\begin{equation}}
\def\eeq{\end{equation}}
\def\beqa{\begin{eqnarray}}
\def\eeqa{\end{eqnarray}}
\def\bet{\begin{tabular}}
\def\eet{\end{tabular}}
\def\bs{\begin{split}}
\def\es{\end{split}}


\def\a{\alpha}  \def\b{\beta}   \def\c{\chi}    
\def\g{\gamma}  \def\G{\Gamma}  \def\e{\epsilon}  
\def\vep{\varepsilon}   \def\tvep{\widetilde{\varepsilon}}
\def\f{\phi}    \def\F{\Phi}  \def\fb{{\ov \phi}}
\def\vf{\varphi}  \def\m{\mu}  \def\mub{\ov \mu}
\def\n{\nu}  \def\nub{\ov \nu}  \def\o{\omega}
\def\O{\Omega}  \def\r{\rho}  \def\k{\kappa}
\def\kab{\ov \kappa}  \def\s{\sigma}
\def\t{\tau}  \def\th{\theta}  \def\sb{\ov\sigma}  \def\S{\Sigma}
\def\l{\lambda}  \def\L{\Lambda}  \def\p{\psi}


\def\cA{{\cal A}} \def\cB{{\cal B}} \def\cC{{\cal C}}
\def\cD{{\cal D}} \def\cE{{\cal E}} \def\cF{{\cal F}}
\def\cG{{\cal G}} \def\cH{{\cal H}} \def\cI{{\cal I}}
\def\cJ{{\cal J}} \def\cK{{\cal K}} \def\cL{{\cal L}}
\def\cM{{\cal M}} \def\cN{{\cal N}} \def\cO{{\cal O}}
\def\cP{{\cal P}} \def\cQ{{\cal Q}} \def\cR{{\cal R}}
\def\cS{{\cal S}} \def\cT{{\cal T}} \def\cU{{\cal U}}
\def\cV{{\cal V}} \def\cW{{\cal W}} \def\cX{{\cal X}}
\def\cY{{\cal Y}} \def\cZ{{\cal Z}}

\def\mC{\mathbb{C}} \def\mP{\mathbb{P}}  
\def\mR{\mathbb{R}} \def\mZ{\mathbb{Z}} 
\def\mT{\mathbb{T}} \def\mN{\mathbb{N}}
\def\mH{\mathbb{H}} \def\mX{\mathbb{X}}

\def\CP{\mathbb{CP}}
\def\RP{\mathbb{RP}}
\def\Z{\mathbb{Z}}
\def\N{\mathbb{N}}
\def\H{\mathbb{H}}

\newcommand{\rmd}{\mathrm{d}}
\newcommand{\rmx}{\mathrm{x}}

\def\tA{ {\widetilde A} } 

\def\one{{\hbox{\kern+.5mm 1\kern-.8mm l}}}
\def\zero{{\hbox{0\kern-1.5mm 0}}}


\newcommand{\bra}[1]{{\langle {#1} |\,}}
\newcommand{\ket}[1]{{\,| {#1} \rangle}}
\newcommand{\braket}[2]{\ensuremath{\langle #1 | #2 \rangle}}
\newcommand{\Braket}[2]{\ensuremath{\langle\, #1 \,|\, #2 \,\rangle}}
\newcommand{\norm}[1]{\ensuremath{\left\| #1 \right\|}}
\def\corr#1{\left\langle \, #1 \, \right\rangle}
\def\vac{|0\rangle}


\def\d{ \partial } 
\def\zb{{\bar z}}

\newcommand{\sq}{\square}
\newcommand{\IP}[2]{\ensuremath{\langle #1 , #2 \rangle}}    

\newcommand{\floor}[1]{\left\lfloor #1 \right\rfloor}
\newcommand{\ceil}[1]{\left\lceil #1 \right\rceil}

\newcommand{\dbyd}[1]{\ensuremath{ \frac{\d}{\d {#1}}}}
\newcommand{\ddbyd}[1]{\ensuremath{ \frac{\d^2}{\d {#1}^2}}}

\newcommand{\Zd}{\ensuremath{ Z^{\dagger}}}
\newcommand{\Xd}{\ensuremath{ X^{\dagger}}}
\newcommand{\Ad}{\ensuremath{ A^{\dagger}}}
\newcommand{\Bd}{\ensuremath{ B^{\dagger}}}
\newcommand{\Ud}{\ensuremath{ U^{\dagger}}}
\newcommand{\Td}{\ensuremath{ T^{\dagger}}}

\newcommand{\T}[3]{\ensuremath{ #1{}^{#2}_{\phantom{#2} \! #3}}}		

\newcommand{\tr}{\operatorname{tr}}
\newcommand{\sech}{\operatorname{sech}}
\newcommand{\Spin}{\operatorname{Spin}}
\newcommand{\Sym}{\operatorname{Sym}}
\newcommand{\Com}{\operatorname{Com}}
\def\adj{\textrm{adj}}
\def\id{\textrm{id}}

\def\ha{\frac{1}{2}}
\def\tha{\tfrac{1}{2}}
\def\wt{\widetilde}
\def\ra{\rangle}
\def\la{\langle}

\def\pb{\ov\psi}
\def\pt{\widetilde{\psi}}
\def\at{\widetilde{\a}}
\def\cb{\ov\chi}
\def\d{\partial}
\def\db{\bar\partial}
\def\delb{\bar\partial}
\def\dbar{\ov\partial}
\def\dag{\dagger}
\def\dalpha{{\dot\alpha}}
\def\dbeta{{\dot\beta}}
\def\dgamma{{\dot\gamma}}
\def\ddelta{{\dot\delta}}
\def\ad{{\dot\alpha}}
\def\bd{{\dot\beta}}
\def\dg{{\dot\gamma}}
\def\dd{{\dot\delta}}
\def\th{\theta}
\def\Th{\Theta}
\def\eb{{\ov \epsilon}}
\def\gb{{\ov \gamma}}
\def\wb{{\ov w}}
\def\Wb{{\ov W}}
\def\D{\Delta}
\def\DD{\Delta^\dag}
\def\Db{\ov D}

\def\ov{\overline}
\def\Slash{\, / \! \! \! \!}
\def\dslash{\partial\!\!\!/} 
\def\Dslash{D\!\!\!\!/\,\,}
\def\fslash#1{\slash\!\!\!#1}
\def\Fslash#1{\slash\!\!\!\!#1}

\def\del{\partial}
\def\delb{\bar\partial}
\newcommand{\ex}[1]{{\rm e}^{#1}} 
\def\ii{{i}}

\newcommand{\vs}[1]{\vspace{#1 mm}}

\newcommand{\ve}{{\vec{\e}}}
\newcommand{\shalf}{\frac{1}{2}}

\newcommand{\lb}{\rangle}
\newcommand{\al}{\ensuremath{\alpha'}}
\newcommand{\ap}{\ensuremath{\alpha'}}

\newcommand{\bean}{\begin{eqnarray*}}
\newcommand{\eean}{\end{eqnarray*}}
\newcommand{\ft}[2]{{\textstyle {\frac{#1}{#2}} }}

\newcommand{\hsp}{\hspace{0.5cm}}
\def\half{{\textstyle{1\over2}}}
\let\ci=\cite \let\re=\ref
\let\se=\section \let\sse=\subsection \let\ssse=\subsubsection

\newcommand{\dpb}{D$p$-brane}
\newcommand{\dpbs}{D$p$-branes}

\def\gh{{\rm gh}}
\def\sgh{{\rm sgh}}
\def\NS{{\rm NS}}
\def\R{{\rm R}}
\def\Qp{Q_{\rm P}}
\def\QP{Q_{\rm P}}

\newcommand\dott[2]{#1 \! \cdot \! #2}

\def\eo{\overline{e}}


\def\p{\partial}
\def\h{{1\over 2}}

\def\d{\partial}
\def\la{\lambda}
\def\eps{\epsilon}
\def\bb{\bigskip}
\def\tg{\widetilde\gamma}
\newcommand{\dm}{\begin{displaymath}}
\newcommand{\edm}{\end{displaymath}}
\renewcommand{\b}{\widetilde{B}}
\newcommand{\gm}{\Gamma}
\newcommand{\ac}[2]{\ensuremath{\{ #1, #2 \}}}
\renewcommand{\ell}{l}
\newcommand{\z}{\ell}
\def\bb{$\bullet$}
\def\Qbar{{\bar Q}_1}
\def\QPbar{{\bar Q}_p}

\def\q{\quad}

\def\bn{B_\circ}

\let\a=\alpha \let\b=\beta \let\g=\gamma 
\let\e=\epsilon
\let\c=\chi \let\th=\theta  \let\k=\kappa
\let\l=\lambda \let\m=\mu \let\n=\nu \let\x=\xi \let\r=\rho
\let\s=\sigma 
\let\vp=\varphi \let\vep=\varepsilon
\let\w=\omega  \let\G=\Gamma \let\D=\Delta \let\Th=\Theta \let\P=\Pi \let\S=\Sigma

\let\t=\tilde

\def\h{{1\over 2}}

\def\r{\rightarrow}
\def\Ri{\Rightarrow}

\def\nn{\nonumber\\}
\let\bm=\bibitem
\def\Kt{{\widetilde K}}
\def\b{\bigskip}

\let\p=\partial

\newcommand{\ma}{m^{}_{\!\!\!\;A}}
\newcommand{\mb}{m^{}_{\!B}}
\newcommand{\mc}{m^{}_{\!\!\;\scalebox{0.6}{$C$}}}

\makeatletter
\def\blfootnote{\xdef\@thefnmark{}\@footnotetext}  
\makeatother


\newcommand{\spectral}{Schwimmer:1986mf}

\newcommand{\sv}{Strominger:1996sh}
\newcommand{\cvet}{Cvetic:1996xz,*Cvetic:1997uw}
\newcommand{\mm}{Balasubramanian:2000rt,*Maldacena:2000dr}
\newcommand{\MM}{Balasubramanian:2000rt,*Maldacena:2000dr}

\newcommand{\lmone}{Lunin:2001ew}
\newcommand{\lmtwo}{Lunin:2001pw}
\newcommand{\lmfour}{Lunin:2001jy}
\newcommand{\lmRot}{Lunin:2001fv}
\newcommand{\lmAdS}{Lunin:2001jy}
\newcommand{\lmm}{Lunin:2002iz}

\newcommand{\GMR}{Gutowski:2003rg}
\newcommand{\CarMcCon}{Cariglia:2004wu}

\newcommand{\lunin}{Lunin:2004uu}
\newcommand{\gmsone}{Giusto:2004id}
\newcommand{\gmstwo}{Giusto:2004ip}

\newcommand{\bw}{Bena:2004de}
\newcommand{\fuzz}{Mathur:2005zp,*Bena:2007kg,*Skenderis:2008qn,*Mathur:2012zp}
\newcommand{\kst}{Kanitscheider:2007wq}
\newcommand{\cern}{Mathur:2009hf}

\newcommand{\mtone}{Mathur:2011gz}
\newcommand{\mttwo}{Mathur:2012tj}
\newcommand{\lmt}{Lunin:2012gp}

\newcommand{\fuzzrefs}{\bw,Giusto:2004kj,Jejjala:2005yu,Bena:2005va,Berglund:2005vb,
Saxena:2005uk,*Balasubramanian:2006gi,Bena:2006kb,*Bena:2007qc,\kst,
Ford:2006yb,*Bena:2010gg,*Bena:2011uw,*Giusto:2011fy,*Giusto:2012gt}

\newcommand{\orbifoldrefs}{Arutyunov:1997gt,*Arutyunov:1997gi,*deBoer:1998ip,*Dijkgraaf:1998gf,
*Seiberg:1999xz,*Larsen:1999uk,*David:1999zb,*Jevicki:1998bm}


\begin{flushright}
\end{flushright}

\begin{center}
{\LARGE D1-D5-P microstates at the cap} \\
\vspace{18mm}
{\bf Stefano Giusto}${}^{1,2}$, ~{\bf Oleg Lunin}${}^{3}$, ~{\bf Samir D. Mathur}${}^{4}$, ~{\bf David Turton}${}^{4}$
\vspace{14mm}

${}^{1}$Dipartimento di Fisica ``Galileo Galilei'',\\
Universit\`a di Padova,\\ Via Marzolo 8, 35131 Padova, Italy\\

\vskip 2mm

${}^{2}$INFN, Sezione di Padova,\\
Via Marzolo 8, 35131, Padova, Italy\\

\vskip 2 mm

${}^{3}$Department of Physics,\\ University at Albany (SUNY),\\ Albany, NY 12222, USA\\ 

\vskip 2 mm

${}^{4}$Department of Physics,\\ The Ohio State University,\\ Columbus,
OH 43210, USA\\ 

\vskip 10 mm

\blfootnote{stefano.giusto@pd.infn.it,~olunin@albany.edu,~mathur.16@osu.edu,~turton.7@osu.edu}

\end{center}

\begin{abstract}
\b 

\noindent

The geometries describing D1-D5-P bound states in string theory have three regions: flat asymptotics, an anti-de Sitter throat, and a `cap' region at the bottom of the throat.
We identify the CFT description of a known class of supersymmetric D1-D5-P microstate geometries which describe degrees of freedom in the cap region.
The class includes both regular solutions and solutions with conical defects, and generalizes configurations with known CFT descriptions: 
a parameter related to spectral flow in the CFT is generalized from integer to fractional values.
We provide strong evidence for this identification by comparing the massless scalar excitation spectrum between gravity and CFT and finding exact agreement.

\end{abstract}

\thispagestyle{empty}

\newpage

\baselineskip=15pt
\parskip=3pt

\section{Introduction}
\label{sec:intro}

\begin{figure}[h!]
\begin{center}
\includegraphics[width=15cm]{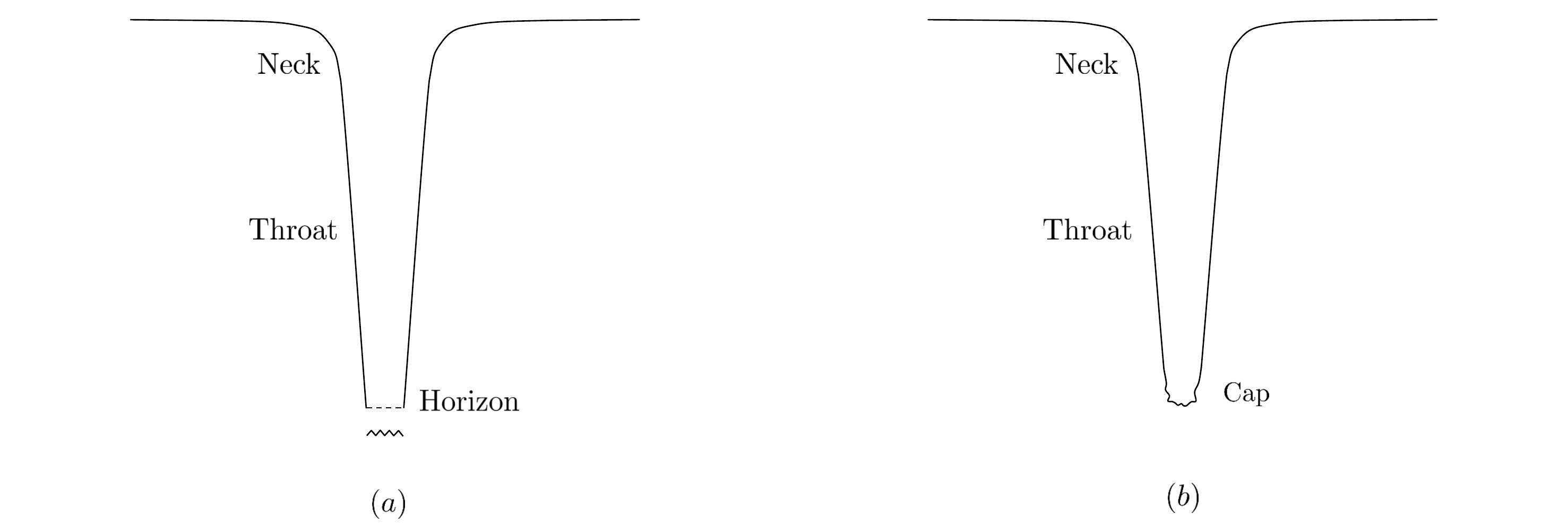}
\caption{(a) Traditional black hole metric with horizon; (b) Fuzzball structure with cap.}
\label{f1}
\end{center}
\end{figure}

A traditional extremal black hole metric involves a region of flat space at infinity, an intermediate `neck' region, an infinite `throat' and a horizon. This structure is depicted in Fig.\;\ref{f1}\,(a). In recent years we have learnt that black hole microstates in string theory have a different structure, depicted in Fig.\;\ref{f1}\,(b). We still have flat space at infinity, the neck, and the throat, but the spacetime ends in a fuzzy `cap' instead of a regular horizon~\cite{\fuzz}. The cap degrees of freedom are inherently quantum in their nature, however one can probe this physics by studying individual microstates which have geometrical caps.

Does this `fuzzball' structure apply to all states of the extremal black hole? 
In the early days of  microstate constructions, it was believed by some that fuzzballs might account for only a fraction of the states of the extremal system, and that the other states would have regular horizons. This belief extended, of course, to nonextremal black holes as well. 
But this scenario would lead back to the information paradox~\cite{Hawking:1974sw,*Hawking:1976ra}.

In Ref.\;\cite{\cern} it was shown that the information paradox cannot be resolved by including small corrections to Hawking's computation. To resolve the paradox, the evolution of low energy modes at the horizon must differ by order unity from the traditionally expected evolution around a smooth horizon. 
This fact suggests that for both extremal and non-extremal black holes, all states must be fuzzballs; no state should have a regular horizon.

\b

\vspace{-5mm}

One can still ask whether the fuzzballs which have been constructed so far\footnote{For an incomplete list of references, see~\cite{\lmfour,\lmm,\lunin,\gmsone,\gmstwo,\fuzzrefs}.} 
are in some sense special, i.e.~whether or not they have properties expected of black hole microstates. In particular, the traditional geometry of Fig.\;\ref{f1}\,(a) may be deformed in two places:

\vspace{-2mm}
\begin{enumerate}[(i)]
	\item At the neck, as shown in Fig.\;\ref{f1}\,(b). 
	\vspace{-2mm}
	\item At the cap, where the traditional geometry had a horizon\footnote{In some of the literature on the counting of black hole microstates, (i) have been termed `hair' and (ii) have been termed `horizon degrees of freedom' (see e.g.~\cite{Dabholkar:2010rm}). We will avoid using the term `hair' for (i), since relativists use `hair' to describe deformations at the {\it horizon}. We thus term (i) the `neck' degrees of freedom, and (ii) the `cap' degrees of freedom. 
}. 
\end{enumerate}
\vspace{-2mm}

\newpage

Given a dual CFT description, one can ask the following questions. Can one characterize these two sets of degrees of freedom as different kinds of states in the dual CFT? And if so, do we have examples of each of these kinds of states in the construction of gravitational microstates?

We work with extremal black holes carrying D1, D5 and P charges \cite{\sv}, and with the dual orbifold CFT~\cite{Maldacena:1997re,\orbifoldrefs}. 
The CFT has a chiral symmetry algebra whose bosonic generators are the Virasoro operators $L_n$, the $SU(2)_R$ currents $J^a_n$ and the U(1) currents ${\cal J}_n^{i}$ of $T^4$ translations.

In recent work we have found that the neck modes correspond to perturbations where a chiral algebra generator 
($L_{-n}, J^a_{-n}, {\cal J}^i_{-n}$ for integer $n$)
is applied to an unexcited D1-D5 state \cite{Mathur:2011gz,Mathur:2012tj,*Lunin:2012gp}. 
But the starting CFT state may be in a twisted sector, which permits the application of
fractional modes, e.g.~$J^a_{-{1\over k}}$ for integer $k$. 
We thus obtain many states which cannot be written in terms of chiral algebra generators applied to a ground state.
In general, we expect that such a state should {\it not} be a neck mode.

In this paper we support this expectation by identifying states of this type which are nontrivial in the cap region.
We present examples of both neck states and cap states and study their properties. 
The CFT states we study are generalizations of spectral flowed states, and involve Fermi seas filled to a level $s/k$.

The dual geometries to spectral flowed states were studied in~\cite{\lunin,\gmsone,\gmstwo}. When written in Bena-Warner form~\cite{\bw}, these geometries arise from harmonic functions with two poles~\cite{Bena:2005va,Berglund:2005vb,Giusto:2004kj}. The geometries contain a spectral flow parameter $\a$ which was originally taken to be an integer $n_L$. However one may also consider fractional values, as noted in \cite{Jejjala:2005yu,Berglund:2005vb}.

In this paper we propose that the above CFT states are described by the geometries with spectral flow parameter $s/k$. Generically, the CFT states of interest are not unique for given charges, so it is not sufficient to compare charges between gravity and CFT. 
Our identification of states is based on the following evidence, which we regard as compelling: 
\vspace{-2mm}
\begin{enumerate}
	\item The geometries with integer spectral flow parameter were related in~\cite{\gmsone,\gmstwo} to CFT states with all component strings treated equally. With this prescription, for general $s$ the CFT state is uniquely specified.
	\vspace{-1mm}
	\item These CFT states arise from spectral flow on the CFT covering space.
	\vspace{-1mm}
	\item To confirm this prescription, we study the spectrum of scalar excitations around the gravity solutions and find exact matching with the CFT.
\end{enumerate}
\vspace{-1mm}

This paper is organized as follows. In Section \ref{sec:CFT} we review the orbifold CFT and the states arising from integer spectral flow. In Section \ref{sec:frac_sp_fl_CFT} we introduce the CFT states with filled Fermi seas. In Section \ref{sec:geometries} we describe the microstate geometries, and we analyze their conical defects in Section~\ref{sec:conical}. In Section \ref{sec:match} we match the scalar excitation spectrum between gravity and CFT. Section~\ref{sec:discuss} discusses our results, and details are presented in Appendices~\ref{app:BW}--\ref{Sec:AppCFT}.

\section{The D1-D5 CFT}
\label{sec:CFT}

We consider type IIB string theory, compactified as
\be
M_{9,1}\rightarrow M_{4,1}\times S^1\times T^4.
\label{compact}
\ee
We wrap $n_1$ D1 branes on $S^1$, and $n_5$ D5 branes on $S^1\times
T^4$. The bound state of these branes is described by a field
theory. We think of the $S^1$ as being large compared to the $T^4$, so
that at low energies we look for excitations only in the direction
$S^1$.  This low energy limit gives a CFT on
the circle $S^1$.

\subsection{The orbifold point}

It has been conjectured that we can move to a point in moduli space
called the `orbifold point' where the CFT is particularly simple
\cite{\orbifoldrefs}. At this orbifold point the CFT is
a 1+1 dimensional sigma model. The 1+1 dimensional base space is
spanned by $(t,y)$, where $t$ is time and $y$ is a coordinate along the $S^1$, with range
\be
0\le y<2\pi R \,.
\ee
The target space of the sigma model is the symmetrized product of
$N=n_1n_5$ copies of $T^4$,
\be
(T^4)^{N}/S_{N},
\ee
with each copy of $T^4$ giving 4 bosonic excitations $X^1, X^2, X^3,
X^4$. It also gives 4 fermionic excitations, which we call $\psi^1,
\psi^2, \psi^3, \psi^4$ for the left movers, and $\bar\psi^1,
\bar\psi^2,\bar\psi^3,\bar\psi^4$ for the right movers. The fermions can be
antiperiodic or periodic around the $\sigma$ circle. If they are
antiperiodic on the $S^1$ we are in the Neveu-Schwarz (NS) sector, and
if they are periodic on the $S^1$ we are in the Ramond (R)
sector.

\subsection{Twisted sectors}

Since we orbifold by the symmetric group $S_{N}$, the theory contains twisted sectors, which can be obtained by acting with twist operators $\sigma_n$ on an untwisted state. Suppose we insert a
twist operator at a point $z$ in the base space. As we circle the
point $z$, different copies of $T^4$ get mapped into each other. Let
us denote the copy number by a subscript $a=1, 2, \dots n$. Every time one circles the twist operator
$
\sigma_{(123\dots n)},
$
the fields $X^i_{(a)}$ get mapped as
\begin{equation}
X^i_{(1)} \rightarrow
X^i_{(2)} \rightarrow
\cdots
\rightarrow
X^i_{(n)} \rightarrow X^i_{(1)},
\end{equation}
and the other copies of $X^i_{(a)}$ are unchanged. We have a similar
action on the fermionic fields. 
Each set of linked copies of the CFT is called
one `component string'.

\subsection{Symmetries}

The CFT has a superconformal ${\cal N}=4$ symmetry in both
the left and right sectors, generated by operators $L_{n}, G^\pm_{r},
J^a_n$ for the left movers and $\bar L_{n}, \bar G^\pm_{r}, \bar
J^a_n$ for the right movers. 

Each $\cN = 4$ algebra has an internal R-symmetry group
SU(2), so there is
a global symmetry group $SU(2)_L\times SU(2)_R$.  We denote the
quantum numbers in these two $SU(2)$ groups as
\be
SU(2)_L: ~(j, m);~~~~~~~SU(2)_R: ~ (\bar j, \bar m).
\ee
In the geometrical setting of the CFT, this symmetry arises from the
rotational symmetry in the 4 space directions of $M_{4,1}$ in
Eq.~\eqref{compact},
\be
SO(4)_E\simeq SU(2)_L\times SU(2)_R.
\ee
Here the subscript $E$ stands for `external', which denotes that these
rotations are in the noncompact directions. These quantum numbers
therefore give the angular momenta of quanta in the gravity
description. 

The fermions can be grouped into representations of $SU(2)_L\times SU(2)_R$. The left fermions give two $j=\h$ representations of $SU(2)_L$ and are singlets of $SU(2)_R$:
\bea\label{FermionsPsi}
\psi^+={1\over \sqrt{2}}(\psi^1+i\psi^2) \,, &&~~\t\psi^+={1\over \sqrt{2}}(\psi^3+i\psi^4) \,, \nn
\psi^-={1\over \sqrt{2}}(\psi^1-i\psi^2) \,, &&~~\t\psi^-={1\over \sqrt{2}}(\psi^3-i\psi^4) \,.
\eea
Similarly, the right fermions $\bar\psi^i$ can be grouped into representations that are doublets of $SU(2)_R$ and singlets of $SU(2)_L$.

\subsection{States in the NS sector}

The vacuum of the CFT is in the NS sector, and is denoted
\be
|0\rangle_{NS}: ~~~h=\bar h=0, ~~~j=m=0, ~~~\bar j=\bar m =0 \,.
\ee
Each of the $N$ copies of the $c=6$ CFT is `singly wound'; i.e., there are no twists. The fermions are antiperiodic around the circles. There are no bosonic or fermionic excitations on any of the circles. The gravity dual of this state is global $AdS_3\times S^3\times T^4$.

Now we consider the chiral primaries of the theory, which correspond to states with charges $m=j=h$ on the left sector and $\bar m =\bar j = \bar h$ on the R sector. (For ease of notation, we shall often use `chiral primary' to refer both to the operators and the corresponding states). The vacuum $|0\rangle_{NS}$ is the lowest dimension chiral primary. 

Other chiral primaries are obtained by twisting together $k$ component strings, to make a component string with winding $k$. On this `multiwound' component string, we can have fractional excitations of bosons and fermions. 
A chiral primary has the highest $SU(2)$ charges for given dimension, so we take no bosonic excitations. We place fermionic excitations $\psi^+, \t\psi^+$ in the lowest allowed levels on the left, respecting the Pauli exclusion principle. We do a similar filling of Fermi levels on the right. The full construction of such chiral primaries is described in \cite{\lmtwo}, where the notation for this chiral primary was $\sigma^{--}_{k}$. The symbol $\sigma$ stands for the twist linking different copies of the $c=6$ CFT, and the subscript $k$ describes the order of the twist. The chiral primaries $\sigma^{--}_k$ have quantum numbers
\be
\sigma^{--}_k:~~~h=j=m={k-1\over 2}, ~~~\bar h = \bar j =\bar m = {k-1\over 2} \,.
\ee
The superscript $--$ says that on each of the left and right sides we have chosen the lower of two possibilities for the spin: we can add a further layer of fermions to any of these sides to get a higher spin chiral primary, so overall we have operators  $\sigma^{\pm\pm}_k$. This gives
\be
\sigma^{++}_k:~~~h=j=m={k+1\over 2}, ~~~\bar h = \bar j =\bar m = {k+1\over 2} \,.
\ee
We can also construct anti-chiral primaries, which have the lowest $m, \bar m$ for their given dimensions:
\be
\t\sigma^{--}_k:~~~h=j={k-1\over 2},~~m=-{k-1\over 2}, ~~~~~\bar h = \bar j  = {k-1\over 2}, ~~\bar m=-{k-1\over 2} \,.
\label{anti}
\ee
To construct these anti-chiral primaries we replace the fermions $\psi^+,\t\psi^+$ by $\psi^-,\t\psi^-$ on the left, and make a similar change on the right. 

\subsection{Spectral flow}

The CFT arising from the D1-D5 brane bound state is in the Ramond (R)
sector, since the periodicities of the
fermions around the $S^1$ are inherited from the behavior of fermionic
supergravity fields around the $S^1$ in \eqref{compact}. One can map states in the NS sector to states in the R sector using spectral flow \cite{\spectral}. Under spectral flow by $\alpha$ units, the dimensions and charges change as follows:
\be
h'=h+\alpha m +\alpha^2 {c\over 24} \,,
\label{hspectral}
\ee
\be
m'=m+\alpha {c\over 12} \,.
\label{mspectral}
\ee
We can perform a similar spectral flow operation for the right movers, with a parameter $\bar \alpha$. 

With $\alpha=1$ we map states in the NS sector to states in the R sector. First consider the NS vacuum state for a single copy of the $c=6$ CFT. This state has $h=m=0$. After spectral flow we obtain 
\be
h'={1\over 4} \,, \qquad m=\h \,.
\label{hone}
\ee
This is the lowest dimension state in the Ramond sector, so we have obtained an R ground state. But we note that this state has nonzero $m$; in fact it is the spin up member of a $j=\h$ multiplet under the group $SU(2)_L$. We will term this spin carried by the component string as the `base spin'. This base spin arises from the fact that the R sector has fermion zero modes, which can be chosen to be in spin up or spin down states. A similar story holds for the right sector, which also has a base spin that can take values $\bar m =\pm \h$. 

We can also get R sector ground states by performing spectral flow with $\alpha=1$ on an antichiral primary (\ref{anti}). Since $k$ copies are twisted together, the total central charge of the involved copies is $c=6k$. We get
\be
h'={k\over 4} \,, \qquad m=\h \,.
\label{hone-1}
\ee
Just as $h={1\over 4}$ in (\ref{hone}) was the lowest energy state in the R sector for one copy of the $c=6$ CFT, we find that ${k\over 4}$ is the lowest energy state in the R sector for $k$ copies of this CFT. The `base spin' of this $k$ twisted component string is again a multiplet with $j=\h, \bar j=\h$, and spectral flow with $\alpha=1, \bar \alpha=1$ from the antichiral primary $\t\sigma^{--}_k$ brings us to the $m=\h, \bar m=\h$ 
state of this multiplet. We call this state $|0^{++}_k\rangle_R$. The state with $m=-\h, \bar m=\h$ is obtained if we start with the antichiral primary $\t\sigma^{+-}_k$, and so on for all four members of the multiplet:
\be
\t\sigma^{--}_k\r |0^{++}_k\rangle_R, ~~\t\sigma^{+-}_k\r |0^{-+}_k\rangle_R, ~~\t\sigma^{-+}_k\r |0^{+-}_k\rangle_R, ~~\t\sigma^{++}_k\r |0^{--}_k\rangle_R
\label{notation}
\ee

\subsection{The 2-charge D1-D5 states}

We now describe the Ramond ground states of the 2-charge D1-D5 system.   We have $N=n_1n_5$ copies of the $c=6$ CFT which can be twisted into component strings with windings $k_1, k_2, \dots$, with $\sum_i k_i=N$. For each component string we can choose the fermion zero modes in different ways. We will consider only the states that arise from the base spins $m=\pm\h, \bar m=\pm \h$; these correspond to all D1-D5 states that do not involve distortions of the $T^4$\;\cite{\lmfour}.\footnote{Including other possibilities for the fermion zero modes gives the remaining states of the theory \cite{\lmm,\kst}.} 
The total central charge is $c=6N$, and the dimension of the R states is 
\be
h=\bar h ={c\over 24}={N\over 4} \,.
\ee
The charges $m, \bar m$ depend on the choices of base spins.

We will be interested in a subclass of states that have $U(1)\times U(1)$ axial symmetry in the gravity description. To get such states we take all component strings to have the same winding $k$ and the same base spin. We must choose $k$ to divide $N=n_1n_5$, giving
\be
n_c={N\over k}
\ee
component strings, each with winding $k$. Suppose we take all base spins to be $-\h$ on both the left and the right sides. We denote this state by $|0^{--}\rangle_R$. Comparing with the notation in (\ref{notation}), we omit the $k$ label when denoting a state of the full CFT rather than a state on a single component string (the label $k$ is understood to be implicit).
The quantum numbers are
\be
|0^{--}\rangle_R:~~~h=\bar h={N\over 4} \,, ~~~~m=-{N\over 2k} \,, ~~\bar m=-{N\over 2k}
\label{start}
\ee
where the charges arise from $m=\bar m=-\h$ on each of the $n_c$ component strings. Similarly, we have
\be
|0^{+-}\rangle_R:~~~h=\bar h={N\over 4} \,, ~~~~m={N\over 2k} \,, ~~\bar m=-{N\over 2k} \,,
\label{startq}
\ee\be
|0^{-+}\rangle_R:~~~h=\bar h={N\over 4} \,, ~~~~m=-{N\over 2k} \,, ~~\bar m={N\over 2k} \,,
\label{startw}
\ee\be
|0^{++}\rangle_R:~~~h=\bar h={N\over 4} \,, ~~~~m={N\over 2k} \,, ~~\bar m={N\over 2k} \,.
\label{starte}
\ee

\subsection{3-charge D1-D5-P states from integer spectral flow} \label{sec:int_sp_fl}

Excited states of the D1-D5 system are obtained by adding bosonic and fermionic excitations to the component strings. To obtain BPS states with momentum charge P, we excite the left movers but leave the right movers in a Ramond ground state. The resulting  momentum charge is
\be
n_p=h-\bar h \,.
\ee
Some of the BPS states at the orbifold point will be lifted when we deform the theory away from the orbifold point, while others will continue to give BPS states at generic points in moduli space. 

A particular subclass of BPS states can be obtained by performing spectral flow on the left movers by an additional amount $\alpha=2n_L$, where $n_L$ is an integer. Such a spectral flow takes an R sector state to the NS sector and back to the R sector, but in the process a R sector ground state ends up as a R sector {\it excited} state. Suppose we start with the state $|0^{-+}\rangle_R$ given in (\ref{startw}). Then spectral flow by $\alpha=2n_L$ gives  the quantum numbers
\be
h'={N\over 4}-{n_LN\over k}+n_L^2 N \,,
\label{hthree}
\ee
\be
m'=-{N\over 2k}+n_L N \,.
\label{htwo}
\ee
We can readily see which excitations on the component strings lead to these quantum numbers. Each component string has twist $k$ and is in the R sector. Thus the fermions have modes that are periodic after $k$ cycles around the $S^1$. 

We can start with acting by  $\psi^+_{0}, \t\psi^+_0$ on each component string; this moves the base spin of the left movers from $-\h$ to $\h$. Next, we can add fermions $\psi^+_{-{1\over k}}, \t\psi^+_{-{1\over k}}$ to each component string. We can follow up with a layer of fermions $\psi^+_{-{2\over k}}, \t\psi^+_{-{2\over k}}$, and so on, adding $s=k n_L$ layers in all. These fermions have $m=\h$ each, so each layer adds $m=1$ to each component string. Thus overall we add $\Delta m=n_c s=n_L N$ units of charge to the charge that we already had from the base spin, and this gives (\ref{htwo}). 

Now we consider the dimension $\Delta h$ contributed by these fermions. The modes $\psi_0, \t\psi_0$ contribute no dimension. The dimension  contributed by the fermions $\psi^+_{-{1\over k}}, \t\psi^+_{-{1\over k}}$ is ${2\over k}$. The dimension contributed by the fermions $\psi^+_{-{2\over k}}, \t\psi^+_{-{2\over k}}$ is ${4\over k}$ and so on. Adding these contributions, and recalling that we have $n_c$ component strings, we find the contribution of these fermions to be
\be
\Delta h=n_c \times 2\times {1\over k} [ 0+1+2+\dots +(s-1)]={N\over k}{(s-1)s\over k}=-{n_LN\over k}+n_L^2 N \,.
\ee
Adding this to the dimension ${N\over 4}$ of the R ground state (\ref{startw}) we get (\ref{hthree}). 
It will be helpful to write out this state explicitly. For $n_L=1$ we have on each component string
\bea
|0^{-+}_k\rangle_{n_L=1}&=&\t\psi^+_{-{ k-1\over k}}\psi^+_{-{k-1\over k}}\dots\t\psi^+_{-{2\over k}}\psi^+_{-{2\over k}}\t\psi^+_{-{1\over k}}\psi^+_{-{1\over k}}\t\psi^+_{0}\psi^+_{0}|0_k^{-+}\rangle_R\nn
&=&\t\psi^+_{-{ k-1\over k}}\psi^+_{-{k-1\over k}}\dots\t\psi^+_{-{2\over k}}\psi^+_{-{2\over k}}\t\psi^+_{-{1\over k}}\psi^+_{-{1\over k}}|0^{++}_k\rangle_R \,.
\label{hfivea}
\eea
where in the second step we used
\be
\t\psi^+_{0}\psi^+_{0}|0^{-+}_k\rangle_R=|0^{++}_k\rangle_R \,.
\ee

The right movers are left unchanged from their configuration in the state (\ref{startw}), so the momentum charge of this state is
\be
n_p=h-\bar h = -{n_LN\over k}+n_L^2 N \,.
\ee

Similarly, we can start with the 2-charge state $|0^{++}\rangle_R$ and perform a spectral flow on the left movers by $2n_L$ units to get a state $|0^{++}\rangle_{n_L}$. The quantum numbers of this state are
\be
h'={N\over 4}+{n_LN\over k}+n_L^2 N, ~~~m'={N\over 2k}+n_L N \,.
\label{hthrees}
\ee
To obtain this state we start with $|0^{++}\rangle_R$, and note that this time we cannot act with the zero modes $\psi^+_0, \t\psi^+_0$ since the base spin is already $\h$. Thus the modes we apply start with $\psi^+_{-{1\over k}}, \t\psi^+_{-{1\over k}}$. We apply $s=kn_L$ such fermion pairs on each component string, and find that we get the quantum numbers (\ref{hthrees}). For $n_L=1$ the state on each component string is
\bea
|0^{++}_k\rangle_{n_L=1}
&=&\t\psi^+_{-1}\psi^+_{-1}\t\psi^+_{-{ k-1\over k}}\psi^+_{-{k-1\over k}}\dots\t\psi^+_{-{2\over k}}\psi^+_{-{2\over k}}\t\psi^+_{-{1\over k}}\psi^+_{-{1\over k}}|0^{++}_k\rangle_R \,.
\label{hfiveas}
\eea
The gravity solutions dual to these states were constructed in \cite{\lunin,\gmsone,\gmstwo}.

\b\b\b

\subsection{Writing states with spectral flow $n_L$ in terms of chiral algebra excitations}

Consider the state (\ref{hfivea}). We now show that this state can be obtained by acting on a Ramond ground state by the {\it integer}-moded chiral algebra generators. We can write
\be
J^+_{-1}~\r~\t\psi^+_{-{1\over k}}\psi^+_{-{k-1\over k}}+\t\psi^+_{-{2\over k}}\psi^+_{-{k-2\over k}}+\dots + \t\psi^+_{-{k-1\over k}}\psi^+_{-{1\over k}}
\label{hfour}
\ee
where we have used the fact that we are acting on a $k$ twisted component string, and kept only negative index modes of $\psi^+, \t\psi^+$ since the positive index modes will annihilate the states that we will act on. First consider the state $J^+_{-1}|0^{++}\rangle_R$. We will get a pair of fermionic excitations, one $\psi^+$ and one $\t\psi^+$. The indices on these fermions can correspond to any fermion pair in (\ref{hfour}). Now consider $(J_{-1}^+)^2|0\rangle_R$. We now have two sets of fermions, and the indices for these must come from two different choices of fermion pairs in (\ref{hfour}), since we cannot act with the same fermion mode twice. Proceeding in this manner, we find that
\be
(J^+_{-1})^{k-1}|0^{++}\rangle_R=(k-1)!|0^{-+}\rangle_{n_L=1}
\label{hten}
\ee
so we have indeed reproduced the state (\ref{hfivea}) by the action of integer moded chiral algebra generators on a 2-charge Ramond ground state. 

Now consider the state (\ref{hfiveas}). This state has an extra layer of fermions, and can be obtained as
\be
J^+_{-2}(J^+_{-1})^{k-1}|0^{++}\rangle_R=(k-1)!|0^{++}\rangle_{n_L=1}
\ee
Proceeding in this manner, we can show that the states $|0^{\pm +}\rangle_{n_L}$ for arbitrary integer $n_L$ can be obtained by acting on $|0^{++}\rangle_R$ with integer-moded operators $J^+_{-n}$. The same construction on the right-movers yields all states $|0^{\pm \pm}\rangle_{n_L}$ for arbitrary integer $n_L$.

\section{A class of CFT states describing cap degrees of freedom} \label{sec:frac_sp_fl_CFT}

Consider the subclass of D1-D5-P states which can be obtained as follows: start with a 2-charge D1-D5 state which is in a R ground state in both left and right sectors, and act on the left sector by generators of the chiral algebra. These 
are the Virasoro generators $L_{-n}$, 
the supercurrents $G^{a\alpha}_{-\n}$, 
and the $SU(2)_R$ generators $J^+_{-n}, J^-_{-n}, J^3_{-n}$.
In addition we have four $U(1)$ currents ${\cal J}^i_{-n}$ coming from the translation symmetries of the torus.

Following general arguments of Brown and Henneaux~\cite{Brown:1986nw}, one expects that the gravity duals of symmetry generators are described  by diffeomorphisms at the boundary of $AdS$ space. Note however that the action of an operator like $L_{-n}$ raises the energy of the state in the CFT, so should do the same in the gravity description.  

It turns out that the gravity dual of a chiral algebra generator is indeed a diffeomorphism at the boundary of the $AdS$ region of the geometry, but it is {\it not} a pure diffeomorphism at the `neck' where the $AdS$ region ends~\cite{\mtone,\mttwo,*\lmt}. The action of the chiral algebra generators can be described as follows: 
We cut off the $AdS$ region of the full geometry, perform a diffeomorphism, and then glue back to flat space (for a depiction, see Fig.~1 of \cite{\mtone}). This process creates a distortion at the `neck', which carries the energy and charge added by the chiral algebra generator. 

In the previous subsection we observed that the D1-D5-P states arising from integer spectral flow can be written in terms of chiral algebra generators on a BPS 2-charge D1-D5 state. The cap structure of a state after integer spectral flow is the same as this ground state; one can say that the distortion created by the P charge arises at the neck. 

We now present a set of states which {\it cannot} be obtained in this way; i.e., they are excitations of a BPS D1-D5 state, but this excitation cannot be obtained by the chiral algebra generators. Thus the corresponding gravity dual is not a `neck' distortion added onto a 2-charge BPS state; the `cap' is genuinely deformed. 
The states arise from a spectral flow transformation in the covering space of the CFT.

\subsection{States arising from `fractional spectral flow'} \label{sec:fr_sp_CFT_sub}

Consider the state $|\Psi\rangle_s$ defined such that the state on each component string takes the form
\bea\label{FrSpFlSt}
|\Psi_k\rangle_s
&=&\t\psi^+_{-{ s\over k}}\psi^+_{-{s\over k}}\dots\t\psi^+_{-{2\over k}}\psi^+_{-{2\over k}}\t\psi^+_{-{1\over k}}\psi^+_{-{1\over k}}|0^{++}_k\rangle_R \,.
\label{hfived}
\eea
The quantum numbers of $|\Psi\rangle_s$ can be obtained by starting with the quantum numbers of $|0^{++}\rangle_R$ and adding the dimensions and charges of the fermions:\footnote{We note in passing that the charges (\ref{httwo}), (\ref{htthree}) do not alone uniquely specify the CFT state. This is demonstrated in Appendix \ref{app:degenerate}. 
}
\be
h={N\over 4} + n_c\times 2 \times {1\over k} [1+2+\dots s]={N\over 4} + {Ns(s+1)\over k^2} \,,
\label{httwo}
\ee
\be
m={n_c\over 2}+n_c s = {N\over 2k} + {Ns\over k} \,.
\label{htthree}
\ee
The state (\ref{FrSpFlSt}) with $s=0$ represents the Ramond vacuum $|0^{++}\rangle_R$, and we have seen that the transition to the state with $s=k$, \eq{hfiveas}, amounts to applying the spectral flow (\ref{hspectral})--(\ref{mspectral}) with $\alpha=2$ to this R vacuum. (For $s=k-1$ we get the state (\ref{hfivea}) which is a spectral flow of $|0^{-+}\rangle_R$.)

For general $s$, the state (\ref{FrSpFlSt}) arises from spectral flow by $2s$ units 
on the $k$--covering space\footnote{Details of the covering space construction may be found in \cite{\lmone,\lmtwo}; see also comment (c) in Appendix \ref{Sec:AppCFT}.}, where ${\tilde h}=kh$ and 
${\tilde c}=\frac{c}{k}$.
On the base space, this operation may be described as `fractional spectral flow'.\footnote{Such a `fractional spectral flow' operation was discussed in~\cite{Avery:2009xr}.}
Comparing the charges \eq{httwo}, \eq{htthree} with the spectral flow transformation (\ref{hspectral}), (\ref{mspectral}) of $|0^{++}_k\rangle_R$, we 
find the effective spectral flow parameter to be
\bea \label{eq:FracSpFl}
\alpha=\frac{2s}{k} \,. 
\eea
Since we are doing spectral flow in the covering space by an even number of units, the periodicities of the CFT fermions in the covering space are unchanged. Thus we may take the CFT fermions to have the same periodicity properties in the state $|\Psi\rangle_s $ as in the original R ground state $|0^{++}\rangle_R$.
On the gravity side, the supergravity fermions are periodic in both cases. 

\subsection{States which cannot be expressed in terms of chiral algebra excitations}

We now demonstrate that for $s$ in the intermediate range 
\be
0 < s < k-1 \,,
\label{range}
\ee
the state $|\Psi\rangle_s$ cannot be obtained by applying integer modes of chiral algebra generators to a 
Ramond ground state. We use proof by contradiction.

\begin{enumerate}[(i)]
\item{Assuming that $|\Psi\rangle_s$ can be constructed by applying superconformal generators to some Ramond vacuum 
$|\Phi\rangle$, and using the fact that these generators do not contain twists, we conclude that the relevant Ramond vacuum should correspond to $n_c=\frac{N}{k}$ component strings with winding $k$.
}
\item{
Using commutation relations of the superconformal algebra, we can order the operators with index $n>0$ to be to the right of operators with index $n\le 0$. The operators with $n>0$ will annihilate the R ground state $|\Phi\rangle$, so we only have operators with index $n\le 0$. Moreover, since R vacua are eigenvectors of $L_0$ and $J^3_0$ and since $J_0^\pm$, $G^\pm_0$ transform vacua into each other, we can assume, without loss of generality, that all operators acting on 
$|\Phi\rangle$ have negative indices.}
\item{
Using the commutation relations, we can further order the operators in the sequence
\be
|\Psi\rangle_s=
\{L_{-n}\}\{ G^{1-}_{-n}\}\{ G^{2-}_{-n}\}\{J^-_{-n}\}\{ J^3_{-n}\} \{G^{1+}_{-n}\}\{ G^{2+}_{-n}\} \{J^+_{-n}\}|\Phi\rangle\,,
\label{class}
\ee
where $\{L_{-n}\}=L_{-n_1}\dots L_{-n_p}$ etc. 
}
\item{The R vacuum $|\Phi\rangle$ discussed in part (i) can contain four types of component strings, $|0^{\pm\pm}_k\rangle$. Let $n_-$ be the number of component strings with left spin down, $|0^{-\pm}_k\rangle$. The left quantum numbers of the state are then
\bea\label{htzeroA}
h_0=\frac{N}{4}\,,\quad m_0=\frac{1}{2}\left(n_c-n_-\right)-\frac{1}{2}n_-=\frac{n_c}{2}-n_-\le \frac{n_c}{2}=\frac{N}{2k}\,.
\eea
}
\item{Comparing (\ref{httwo}) and (\ref{htthree}) with (\ref{htzeroA}), we conclude that the elements of the chiral algebra in (\ref{class}) make the following contributions to the charges:
\bea
\Delta h=h-h_0=\frac{Ns(s+1)}{k^2},\quad \Delta m=m-m_0\ge \frac{Ns}{k} \,.
\eea
In particular, for $s$ in the range (\ref{range}) we find
\bea\label{TheIneq}
\Delta h-\Delta m\le \frac{Ns(s+1-k)}{k}< 0.
\eea
We will now demonstrate that this inequality cannot be satisfied by the generators of the chiral algebra appearing in (\ref{class}).
} 
\item{For the generators appearing in (\ref{class}), we find
\bea
J^\pm_{-n}:\ \Delta h-\Delta m=n\mp 1,\qquad&& J^3_{-n}:\
\Delta h-\Delta m=n,\nonumber\\
G^\pm_{-n}:\ \Delta h-\Delta m=n\mp \frac{1}{2},\qquad&& 
L_{-n}:\
\Delta h-\Delta m=n.\nonumber
\eea
In particular, all these ingredients have $\Delta h-\Delta m\ge 0$, so it is impossible to satisfy the inequality (\ref{TheIneq}). This contradicts (\ref{class}), thus the state $|\Psi\rangle_s$ cannot be obtained by applying integer modes of chiral algebra generators to a 
Ramond ground state. 
}
\end{enumerate}

This argument can be extended to show that the state $|\Psi\rangle_s$ with 
\bea
s=n_Lk+s' \,, \qquad n_L \in \mZ \,, \qquad 0 < s' < k-1 \,
\label{htone}
\eea
cannot be obtained by applying the generators of the superconformal algebra to a Ramond ground state. We again argue by contradiction. Assuming the opposite and writing 
$|\Psi\rangle_s$ as (\ref{class}), we can apply  $(-2n_L)$ units of  spectral flow to this state. This operation transforms the generators  of the superconformal algebra into each other, the R vacuum $|\Phi\rangle$ into a sequence of Virasoro generators acting on some other R vacuum $|\Phi'\rangle$, and the state $|\Psi\rangle_s$ into a state with $\frac{N}{k}$ component strings with length $k$ and charges\footnote{See transformation rules (\ref{hspectral}), (\ref{mspectral}).}
\bea\label{AAcharges}
h'={N\over 4} + {Ns'(s'+1)\over k^2},\qquad 
m'={N\over 2k} + {Ns'\over k} \,.
\eea
Our current assumption implies that the state with charges (\ref{AAcharges}) can be written in the form (\ref{class}) with 
$|\Phi'\rangle$ instead of $|\Phi\rangle$, and this assertion was 
falsified above. 

For later use, we note that since integer spectral flow by $(-2n_L)$ units acts independently on each copy of the CFT, the state $|\Psi\rangle_s$ transforms exactly into the state $|\Psi\rangle_{s'}$.

\subsection{Constraints on $s$} \label{sec:constraints_s}

Given integer parameters $n_1,n_5$ and $k$, let us ask which values of $s$ are allowed\footnote{We always take $n_c = n_1n_5/k$ to be an integer for the states in this paper.}. 
In the state (\ref{hfived}) we find that on each component string the momentum is 
\be
p_c~=~h-\bar h ~= ~2\times {1\over k} [1+2+\dots s]~=~{s(s+1)\over k} \,.
\ee
It has often been assumed that the rules of the orbifold CFT require that this quantity is an integer; this assumption constrains the possible values of $s$ that we can choose. But it turns out that if we start with a D1-D5-P state that satisfies the constraint of integral momentum per component string, and then perform dualities to interchange the D1 charge with the P charge, then the new state does {\it not} satisfy the integrality constraint~\cite{Giusto:2004ip}. We can trace this failure to the following fact. In the orbifold theory one usually assumes that 
$n_1$ and $n_5$ are coprime. But after dualities these charges can have common factors.
To reconcile these facts, we make the following proposal. Writing
\bea
d &=& \gcd(n_1,n_5)\,,
\eea
we propose that in general the momentum per component string can be 
\bea \label{eq:mom-units-d}
h-\bar h &=& {n\over d} \,, \qquad\quad  n \in \mZ \,.
\eea
This condition, which is a slight extension of the usual rules of the orbifold CFT, turns out to be duality covariant; further, it agrees with integrality conditions that are found from the dual gravitational solutions (see Eq.~\eq{eq:BWintegers} and the following discussion).

\section{The dual geometries} \label{sec:geometries}

We have seen that from the CFT perspective, the fractional spectral flow states \eq{FrSpFlSt} are qualitatively different from integer spectral flow states, as the former cannot be obtained by acting with integer chiral algebra generators on Ramond ground states. On the gravity side we thus expect fractional spectral flow states to describe genuine cap degrees of freedom. In this section we identify the gravity description of these states.

The geometries dual to integer spectral flow states like (\ref{hfivea}) and (\ref{hfiveas}) have a $U(1)\times U(1)$ axial symmetry. This reflects the symmetric nature of the state, where all the component strings are equal and have the same base spins and the same fermionic excitations.  The geometries were constructed explicitly in  \cite{\gmsone,\gmstwo}, by taking extremal limits in the general family of rotating 3-charge solutions of \cite{\cvet} and in \cite{\lunin}, by solving the appropriate system of supergravity equations~\cite{Gutowski:2003rg}. 

A systematic technique to construct microstate geometries with $U(1)\times U(1)$ axial symmetry was later developed in \cite{\bw,Bena:2005va,Berglund:2005vb}: the most general axially symmetric microstate is generated by assembling harmonic functions on $\mathbb{R}^3$, subject to some constraints. It was recognized that the geometries of \cite{\gmsone,\gmstwo,\lunin}  represent a subclass of the solutions of \cite{\bw} that are based on harmonic functions with two poles. 

However, the integer spectral flow geometries do not exhaust the class of two-center microstate solutions. This fact was noted in~\cite{Berglund:2005vb} and is shown in Appendix \ref{app2centers} for completeness.
The geometries have the same form as the solutions in \cite{\gmsone,\gmstwo}, but the angular momentum parameters, denoted as $\gamma_1$ and $\gamma_2$ in the following, now take fractional values. The existence of geometries with these values of the parameters was previously noted in~\cite{Jejjala:2005yu}.

It has been an open problem until now to identify the CFT states the remaining two-center geometries are dual to. In this paper we propose that the duals are the fractional spectral flow states (\ref{FrSpFlSt}).

As explained above, we work in type IIB string theory compactified on $M_{4,1}\times S^1\times T^4$, where we denote by $y$ the coordinate of $S^1$ and by $z_a$ ($a=1,\ldots,4$) those of $T^4$. The radius of $S^1$ is $R$ and the volume of $T^4$ is $(2\pi)^4\,V$.
The string metric, RR 2-form and dilaton are
\bea
\label{stringmetric}
ds^2 &=& -\frac{1}{h}\,(dt^2-dy^2) +\frac{Q_p}{h\,f}\,(dt-dy)^2 + h\,f\,\Bigl(\frac{dr^2}{r^2 + a^2\,(\gamma_1+\gamma_2)^2\,\eta}+d\theta^2\Bigr)\nonumber\\
&&+ h\,\Bigl(r^2 + a^2\,\gamma_1\,(\gamma_1+\gamma_2)\,\eta - \frac{Q_1 Q_5\,a^2\,(\gamma_1^2-\gamma_2^2)\,\eta\,\cos^2\theta}{h^2\,f^2}\Bigr)\,\cos^2\theta\,d\psi^2\nonumber\\
&&+ h\,\Bigl(r^2 +a^2\, \gamma_2\,(\gamma_1+\gamma_2)\,\eta + \frac{Q_1 Q_5\,a^2\,(\gamma_1^2-\gamma_2^2)\,\eta\,\sin^2\theta}{h^2\,f^2}\Bigr)\,\sin^2\theta\,d\phi^2\nonumber\\
&&+\frac{Q_p\,a^2\,(\gamma_1+\gamma_2)^2\,\eta^2}{h\,f}\,(\cos^2\theta \,d\psi + \sin^2\theta\,d\phi)^2\nonumber\\
&&-\frac{2\,\sqrt{Q_1 Q_5}\,a}{h\,f}\,(\gamma_1\,\cos^2\theta\,d\psi + \gamma_2\,\sin^2\theta\,d\phi)\,(dt-dy)\nonumber\\
&&-\frac{2\,\sqrt{Q_1 Q_5}\,a\,(\gamma_1+\gamma_2)\,\eta}{h\,f}\,(\cos^2\theta\,d\psi + \sin^2\theta\,d\phi)\,dy + \sqrt{\frac{H_1}{H_5}}\,\sum_{a=1}^4 dz_a^2\,,
\eea
\bea
C^{(2)} &=& - \frac{\sqrt{Q_1 Q_5}\,a\,\cos^2\theta}{H_1\,f}\,(\gamma_2\,dt+\gamma_1\,dy)\wedge d\psi - \frac{\sqrt{Q_1 Q_5}\,a\,\sin^2\theta}{H_1\,f}\,(\gamma_1\,dt+\gamma_2\,dy)\wedge d\phi\nonumber\\
&& + \frac{(\gamma_1+\gamma_2)\,a\,\eta\,Q_p}{\sqrt{H_1 \,f}}\,(Q_1\,dt+Q_5\,dy)\wedge (\cos^2\theta \,d\psi + \sin^2\theta\,d\phi)\nonumber\\
&&-\frac{Q_1}{H_1\,f}\,dt\wedge dy- \frac{Q_5\,\cos^2\theta}{H_1\,f}\,(r^2 +a^2\, \gamma_2 \,(\gamma_1+\gamma_2)\,\eta + Q_1)\,d\psi\wedge d\phi\,,
\eea
\be
e^{2\Phi} = \frac{H_1}{H_5}\,,
\ee
where
\bea
&&a=\frac{\sqrt{Q_1\,Q_5}}{R}\,,\quad Q_p = -a^2\,\gamma_1\,\gamma_2\,\quad \eta = \frac{Q_1\, Q_5}{Q_1\,Q_5 + Q_1\, Q_p + Q_5 \,Q_p}\,,\nonumber\\
&&f = r^2 + a^2\,(\gamma_1+\gamma_2)\,\eta\,(\gamma_1\,\sin^2\theta + \gamma_2\,\cos^2\theta)\,,\nonumber\\
&&H_1=1+\frac{Q_1}{f}\,,\quad H_5=1+\frac{Q_5}{f}\,,\quad h =\sqrt{H_1\,H_5}\,.
\eea
In the duals of the states (\ref{FrSpFlSt}), the value of the parameters $\gamma_1$ and $\gamma_2$ is set to
\be \label{eq:g1g2}
\gamma_1 = -\frac{s}{k}\,,\quad \gamma_2 =\frac{s+1}{k}\,.
\ee
We also introduce for later use
\be \label{eq:gamma}
\gamma ~\equiv~ \gamma_1 + \gamma_2 ~=~ \frac{1}{k}\,.
\ee
The geometry carries D1, D5 charges whose integer values, $n_1$, $n_5$ are related to the dimensionful parameters that appear in the metric by
\be
\label{Q1Q5}
Q_1= \frac{g\,\alpha'^3}{V}\,n_1\,,\quad Q_5 =g\,\alpha'\,n_5\,,
\ee
and a momentum charge $n_p$, given by
\be
\label{Qp}
n_p =  \frac{V\,R^2}{g^2\,\alpha'^4}\,Q_p = \frac{s\,(s+1)}{k}\,\frac{n_1\,n_5}{k}\,.
\ee
The angular momenta are
\bea
\label{JLJR}
J_L &\equiv& \frac{1}{2}\,(J_\phi - J_\psi) =\frac{V}{2\,g^2\,\alpha'}\,Q_1 Q_5\,(\gamma_2-\gamma_1) = \Bigl(s+\frac{1}{2}\Bigr)\,\frac{n_1\,n_5}{k} \,,\nonumber\\
J_R &\equiv& \frac{1}{2}\,(J_\phi+J_\psi)=\frac{V}{2\,g^2\,\alpha'}\, Q_1 Q_5\,(\gamma_2+\gamma_1) =\frac{1}{2}\, \frac{n_1\, n_5}{k}\,. 
\eea
We note that the values of the momentum charge and of the angular momenta are in agreement with the CFT result (\ref{httwo}--\ref{htthree}), using the identifications $n_p = h - \bar h$, $J_L = m$, $J_R = \bar m$.

\addtolength{\textheight}{-0.2 in}
 
\section{Conical defects} \label{sec:conical}

The three-charge geometry (\ref{stringmetric}) is regular at non-vanishing $r$, but for generic values of $s$ and $k$ it has conical singularities at $r=0$. For particular values of $s$ and $k$ the geometry is completely smooth (as noted in~\cite{Jejjala:2005yu}). In this section we give a complete analysis of the conical singularities which arise.

The CFT description developed in Section \ref{sec:CFT} is relevant when the geometry (\ref{stringmetric}) has a long AdS `throat', which exists if $a\ll (Q_1Q_5)^{1/4}$. We begin with analyzing singularities in this regime of parameters, and we comment on the general case at the end of the section. The long AdS region exists when
\bea\label{NearHorPar}
a ~\ll~ (Q_1Q_5)^{1/4}\qquad\Rightarrow\qquad Q_p~\ll~\sqrt{Q_1Q_5} \,,\qquad \eta~\approx ~1 \,.
\eea 
In the near horizon limit, $r\ll \sqrt{Q_1Q_5}$, the metric (\ref{stringmetric}) simplifies\footnote{In this section we focus on the six--dimensional part of the metric.}: 
\bea\label{MaybeCut}
ds_6^2&=&-\frac{f}{\sqrt{Q_1Q_5}}(dt^2-dy^2)+\frac{Q_p}{\sqrt{Q_1Q_5}}
[dt-dy]^2
\nonumber\\
&&{}+\sqrt{Q_1Q_5}\left[\frac{dr^2}{r^2+a^2(\gamma_1+\gamma_2)^2}+d\theta^2+\cos^2\theta d\psi^2+\sin^2\theta d\phi^2\right]\\
&&{}-2a(\gamma_1\cos^2\theta d\psi+\gamma_2\sin^2\theta d\phi)(dt-dy)-
2a(\gamma_1+\gamma_2)(\cos^2\theta d\psi+\sin^2\theta d\phi)dy\,.\nonumber
\eea
The change of variables
\bea\label{FracSPShift-1}
{\tilde t}=\frac{t}{R}, \quad {\tilde y}=\frac{y}{R}, \quad {\tilde r}=\frac{r}{a}, \quad
{\tilde\psi}=\psi-\g_1\frac{t}{R}-\g_2\frac{y}{R},\quad {\tilde\phi}=\phi-\g_2\frac{t}{R}+\g_1\frac{y}{R}
\eea
brings the metric to local AdS$_3\times$S$^3$ form:
\bea\label{SpecFlMtr}
ds_6^2&=&\sqrt{Q_1Q_5}\left[-({\tilde r}^2+\gamma^2)dt^2
+\frac{d{\tilde r}^2}{{\tilde r}^2+\gamma^2}+{\tilde r}^2 d{\tilde y}^2
+d\theta^2+\cos^2\theta d{\tilde\psi}^2+
\sin^2\theta d{\tilde \phi}^2\right].\quad
\eea
Since $\g_1$ and $\g_2$ take fractional values \eq{eq:g1g2}, the change of variables from $(\psi,\phi)$ to $({\tilde\psi},{\tilde\phi})$, is
\bea\label{FracSPShift}
{\tilde\psi}=\psi-\frac{y}{Rk}+\frac{s}{Rk}(t-y),\qquad {\tilde\phi}=\phi-\frac{t}{Rk}-\frac{s}{kR}(t-y) \,.\quad
\eea
This coordinate transformation is often described as `spectral flow', because of its relation to the corresponding operation in the dual CFT. Here we have a spectral flow coordinate transformation with a fractional parameter $s/k$. This `fractional spectral flow' fits nicely with the CFT discussion in Section~\ref{sec:fr_sp_CFT_sub}.

While the metric (\ref{SpecFlMtr}) is locally 
AdS$_3\times$S$^3$, the global identifications induced by (\ref{FracSPShift}) may lead to conical defects, and now 
we discuss their properties.
 
The singularities arise at $r=0$, where the coordinate $y$ is ill-defined, and the change of variables (\ref{FracSPShift}) is not a genuine diffeomorphism. It is convenient to go to the  covering space in coordinates $(y,{\tilde\psi},\tilde\phi)$, then
periodicities of $y$, $\psi$ and $\phi$ translate into identifications for the shifted variables:
\bea\label{OrigLattice}
A:&& ({\tilde y},{\tilde\psi},\tilde\phi)\rightarrow ({\tilde y},{\tilde\psi},\tilde\phi)+2\pi\left(1,-\frac{s+1}{k},\frac{s}{k}\right),\nonumber\\
B:&& ({\tilde y},{\tilde\psi},\tilde\phi)\rightarrow ({\tilde y},{\tilde\psi},\tilde\phi)+2\pi(0,1,0),\\
C:&& ({\tilde y},{\tilde\psi},\tilde\phi)\rightarrow ({\tilde y},{\tilde\psi},\tilde\phi)+2\pi(0,0,1).\nonumber
\eea
A conical singularity can only occur at points which remain fixed under some combination of generators (\ref{OrigLattice}):
\bea\label{ABCgen}
&& A^{\ma}  B^{\mb} C^{\mc}  \,, \qquad\qquad m_I \in \mZ \,.
\eea
If $\ma=0$, this operation does not involve ${\tilde y}$. Then the fixed points occur at $\theta=0$ (if $\mb=0$) or at $\theta=\frac{\pi}{2}$ (if $\mc=0$), and the shifts (\ref{ABCgen}) are consistent with the periodicities of ${\tilde\psi}$ and ${\tilde\phi}$ which guarantee regularity of the metric (\ref{SpecFlMtr}) at these points. 

For $\ma \ne 0$, a fixed point can only occur at $r=0$, since generators $B$ and $C$ do not affect the coordinate ${\tilde y}$. 
There are several possibilities: \nopagebreak

\vspace{2mm}

\noindent
{\underline {\bf Case 1: \,$\gcd(k, s) = \gcd(k, s+1) = 1$}}

If $(k, s, s + 1)$ are pairwise coprime, then the geometry is completely regular.

To see this, we first consider the possibility that a fixed point happens at $\theta\ne 0$. For this to happen, ${\tilde\phi}$ must remain invariant under (\ref{ABCgen}). 
This implies that $\mc = -\frac{s\,\ma}{k} $, which requires $\frac{\ma}{k}$ to be an integer. Then writing $\ma = k m'$, (\ref{ABCgen}) yields the identification
\bea
{\tilde y}\rightarrow {\tilde y}+2\pi k m',\quad 
{\tilde\psi}\rightarrow {\tilde\psi}+ 2\pi\left[\mb-(s+1)m'\right].
\eea
Setting $\mb = (s+1)m'$ gives the identification
\bea
{\tilde y}\rightarrow {\tilde y}+2\pi k m'
\eea
which is the correct periodicity to guarantee regularity of the metric (\ref{SpecFlMtr}).
The potential fixed points at $\theta= 0$ may be analyzed 
in the same way by interchanging ${\tilde\psi}$ and ${\tilde\phi}$.

\vspace{2mm}

\noindent
{\underline {\bf Case 2: \,$\gcd(k,s) >1$, \,$\gcd(k,s+1) =1$}}

In this case, defining $l_1 \equiv \gcd(k,s)$, there is a $\mZ_{l_1}$ orbifold at $(r=0,\theta=\frac{\pi}{2})$.

To see this, write $k=l_1 \hat{k}$, $s=l_1 \hat{s}$ and note that fixed points are obtained
by setting $\ma={\hat k} m'$, $\mc=-{\hat s}m'$, $\mb=0$.
Then the metric and identifications near $(r=0,\theta=\frac{\pi}{2})$ become
\bea\label{ConicSing}
ds_{sing}^2&\approx&\sqrt{Q_1Q_5}
\left[-\frac{dt^2}{k^2}+k^2 \left[ dr^2+r^2\left(\frac{d{\tilde y}}{k}\right)^2\right]
+d\theta^2+\cos^2\theta d{\tilde\psi}^2\right],
\nonumber\\
&&\frac{{\tilde y}}{k}\sim \frac{{\tilde y}}{k}+2\pi \frac{m'}{l_1},\quad {\tilde\psi}\sim{\tilde\psi}-
\frac{2\pi m'}{l_1}(s+1) \,.\nonumber
\eea 

\vspace{2mm}

\noindent
{\underline {\bf Case 3:\, $\gcd(k,s) =1$, \,$\gcd(k,s+1) >1$}}

In this case, defining $l_2 \equiv \gcd(k,s+1)$,
there is a $\mZ_{l_2}$ orbifold at $(r=0,\theta=0)$. 

Writing $k=l_2 \hat{k}$, $s+1=l_2 \hat{t}$, the fixed points are obtained by setting $\ma={\hat k} m'$, $\mb=-{\hat t}m'$. 

\vspace{3mm}

\noindent
{\underline {\bf Case 4:\, $\gcd(k,s) >1$, \,$\gcd(k,s+1) >1$}}

In this case, there is a $\mZ_{l_1}$ orbifold at $(r=0, \theta=\frac{\pi}{2})$ and a $\mZ_{l_2}$ orbifold at $(r=0,\theta=0)$. 

Writing $k=l_1 l_2 \hat{k}$, \,$s=l_1 \hat{s}$, \,$s+1=l_2 \hat{t}$, the fixed points are obtained by setting $\ma=l_1 {\hat k} m'$ and $\ma=l_2{\hat k} m'$. \\

We will now interpret these results from the point of view of the dual CFT. The solutions (\ref{stringmetric}) with $s=0$ have only D1--D5 charges, and correspond to Ramond vacua in the dual CFT. 
These solutions have a $\mZ_k$ orbifold at $(r=0,\theta=\frac{\pi}{2})$.

When $s=nk$ or $s=nk-1$ for integer $n$, the solutions are dual to the CFT states obtained via spectral flow, discussed in Section \ref{sec:int_sp_fl}. These solutions have $\mZ_k$ orbifolds at $(r=0,\theta=\frac{\pi}{2})$ and $(r=0,\theta=0)$ respectively.

For general $s$, we propose that these solutions are dual to the CFT states~\eq{FrSpFlSt}.
We have seen that at particular values of $s$, the orbifold singularity is softer, 
and the geometry is completely smooth if $(k,s,s+1)$ are mutually prime.
A comment on when this is possible is in order. If we take $n_1$ and $n_5$ coprime, then the Bena-Warner conditions for regularity require (see \eq{eq:cond1}-\eq{eq:pc})
\bea \label{eq:pc-1}
p_c ~=~ \frac{s(s+1)}{k} \quad \in ~\mZ \,.
\eea
This means that we are in Case 4 of the above orbifold analysis.

When $n_1$ and $n_5$ are not coprime, $p_c$ does not have to be an integer, which can be seen by doing dualities to interchange the charges, as discussed in Section \ref{sec:constraints_s}. 
In this case, it is not immediately obvious whether the configurations describe bound states. 

Starting with non-coprime $n_1,n_5,n_p$ and interchanging charges by dualities, 
one clearly obtains a resulting configuration in which the new $n_1,n_5,n_p$ are still non-coprime.
However, for example when $\gcd(n_5,n_p)=1$ and $\gcd(n_1,n_5)>1$, permuting $n_1$ and $n_p$
maps a configuration with $n_1, n_5$ non-coprime to a configuration
with $n_1, n_5$ coprime. 
It is straightforward to generate explicit examples of configurations in Case 4 of the orbifold analysis which are U-dual to configurations falling in each of Cases 1--3 of the above orbifold analysis.
It is to be expected that dualities map bound states to bound states, so in this context it seems that all of Cases 1-4 above are of physical relevance.

A further comment is in order. Consider the CFT state
\bea
s=n_Lk+s' \,, \qquad n_L \in \mZ \,, \qquad 0 < s' < k-1 \,.
\label{htone-1}
\eea
Note that the common divisor structure is the same whether we use $s$ or $s'$, and hence the orbifold structure of the cap is the same in both cases. 
In the CFT, the states with parameters $s$ and $s'$ are related by spectral flow by $\a=2n_L$.
So we see that acting on all states in this category, spectral flow by an even integer preserves the cap structure -- even though the resulting state is very different. Since this operation can be written terms of chiral algebra generators, this observation further supports the relation between the chiral algebra generators and degrees of freedom at the neck described in Section \ref{sec:frac_sp_fl_CFT}.

We conclude this section by relaxing the requirement (\ref{NearHorPar}), which was introduced to have a direct connection with CFT and to make formulas more explicit. The shifted coordinates (\ref{FracSPShift}) can be introduced in the general solution (\ref{stringmetric}), and the fixed points of (\ref{ABCgen}) can be analyzed in the resulting geometry. As before, we will only need the behavior of the metric near 
$(r=0, \theta=0)$:
\bea
g_{{\tilde y}{\tilde y}}\approx\frac{g_{rr}}{k^2},\quad 
g_{{\tilde\phi}{\tilde\phi}}\approx\theta^2g_{\theta\theta},
\quad g_{{\tilde\phi}\mu}\sim \theta^2,\quad
g_{{\tilde y}\mu}\sim r^2,
\eea
and near $(r=0, \theta=\frac{\pi}{2})$:
\bea
g_{{\tilde y}{\tilde y}}\approx\frac{g_{rr}}{k^2},\quad 
g_{{\tilde\psi}{\tilde\psi}}\approx
(\theta-\frac{\pi}{2})^2g_{\theta\theta},
\quad g_{{\tilde\psi}\mu}\sim (\theta-\frac{\pi}{2})^2,\quad
g_{{\tilde y}\mu}\sim r^2.
\eea
These are the only properties which were used in the analysis following equation (\ref{ABCgen}), so the properties of the conical defects discussed above hold for the metrics (\ref{stringmetric}) for general parameter ranges, without taking the limit \eq{NearHorPar}.

\b\b

\section{The spectrum of excitations}
\label{sec:match}

We have considered a set of states in the CFT and described gravitational solutions that have the same quantum numbers. To strengthen the identification between the CFT states and the gravity solutions we study the excitation spectrum around the state in each description\footnote{For related calculations involving massless scalar emission from orbifolded JMaRT solutions \cite{Jejjala:2005yu}, see \cite{Chowdhury:2008bd,Avery:2009xr}.}. We will find that these spectra agree, thus providing strong support to our identification.

\subsection{Scalar perturbations around the gravitational solutions}

In this section we consider small perturbations of the supergravity fields
around the solutions \eq{stringmetric}. In particular we consider a graviton $h_{ij}$
or a gauge field quantum $C^{(2)}_{ij}$ with indices on the $T^4$. Such quanta give rise to minimal scalars $\phi$ in the 6-d metric obtained by dimensional reduction on the torus.  Thus the spectrum of excitations is given by solutions of the wave equation
\be
\square \Phi=0 \,.
\label{box}
\ee
We take the ansatz
\be
\Phi=\Phi(t,y, r,\theta,\psi,\phi)=\exp(-i\omega \frac{t}{R}+i\lambda\frac{y}{R}+ip\psi+iq\phi) 
H(r)\Theta(\theta) \,.
\ee
For our metrics (\ref{stringmetric}), the geometry near the cap is given by (\ref{SpecFlMtr}). In terms of the variables \eq{FracSPShift-1} used in the cap geometry, we write
\be
\Phi=\Phi(\tilde t,\tilde y, \tilde r, \theta, \tilde \psi, \tilde \phi)=\exp(-i\tilde\omega \tilde t+i\tilde\lambda \tilde y+ip\tilde\psi+iq\tilde\phi)
\tilde H(\tilde r)\Theta(\theta) \,.
\ee
The radial part of the wave equation (\ref{box}) becomes \cite{Lunin:2002fw}
\be
\frac{1}{\tilde r}\frac{d}{d\tilde r}\left(\tilde r({\tilde r}^2+\gamma^2)\frac{d\tilde H}{d\tilde r}\right)+
\left\{\frac{{\tilde\omega}^2}{{\tilde r}^2+\gamma^2}-
\frac{{\tilde\lambda}^2}{{\tilde r}^2}\right\}\tilde H-\Lambda \tilde H=0 \,.
\ee
Let us now review the solution of this equation, obtained in~\cite{Lunin:2002fw}.
We assume $\tilde\lambda\ge 0$; the equation involves only ${\tilde\lambda}^2$, so we can recover the results for $\tilde\lambda\le 0$ later. Then, defining $x={\tilde r}^2$, the solution regular at $\tilde r=0$ is
\be
H(x)=x^{\tilde\lambda\over 2\gamma}(x+\gamma^2)^{\tilde\omega\over 2\gamma}F[a, a+l+1, c, -{x\over \gamma^2}]
\ee
where 
\be
a=-{l\over 2}+{\tilde\omega+\tilde\lambda\over 2\gamma}, ~~~c=1+{\tilde\lambda\over \gamma} \,.
\ee
To select the solution which dies at large $r$, we use the identity
\bea
F[a,b,c,z]&=&{\Gamma[c]\Gamma[b-a]\over \Gamma[b]\Gamma[c-a]}(-1)^az^{-a}F[a,a+1-c,a+1-b,{1\over z}]\nonumber\\
&&+{\Gamma[c]\Gamma[a-b]\over \Gamma[a]\Gamma[c-b]}(-1)^bz^{-b}F[b,b+1-c,b+1-a,{1\over z}] \,.
\eea
The second term arising this way behaves as $z^{-{l\over 2}-1}$ for large $z$, so it is convergent at large $r$. The first behaves as $z^{l\over 2}$, so its coefficient will have to vanish. Thus either $\Gamma[b]$ has a pole or $\Gamma[c-a]$ has a pole. 
The first possibility gives $b=-n$ with $n\ge 0$, which implies
\be
{(\tilde\omega+\tilde\lambda)\over 2\gamma}=-{l\over 2}-1-n, ~~~n\ge 0 \,.
\ee
Since we have taken $\tilde\lambda\ge 0$, this implies $\tilde\omega <0$, and so we discard this case. 

The second possibility gives $c-a=-n$ with $n\ge 0$, which implies
\be
{\tilde\omega-\tilde\lambda\over 2\gamma}={l\over 2} +1+n, ~~~n\ge 0
\ee
and this determines the allowed spectrum. Noting that a similar result holds for $\tilde\lambda\le 0$, we get the full spectrum as
\be
k\tilde\omega=l+2(n+1)+|k\tilde\lambda|, ~~~n\ge 0
\ee
where we have substituted $\gamma={1\over k}$. Finally, we use the coordinate transformation 
(\ref{FracSPShift-1}) to map back to the original coordinates of the metric (\ref{stringmetric}). This gives
\be
\omega~=~\tilde\omega+(p\gamma_1+q \gamma_2) \,, \qquad\quad
\lambda~=~\tilde\lambda-(p\gamma_2+q \gamma_1) \,.
\ee
Thus the spectrum is
\bea
k\omega &=&({l}+2(n+1))+|k\lambda+(p(s+1)-q  s)|+(-ps+q (s+1))\,.
\label{relation1}
\eea
It will be convenient to express this spectrum in terms of the left and right angular momenta $m$ and $\bar m$, via
\be
p~=~{\bar m}-m, ~~~q~=~{\bar m}+m \,.
\ee
This gives the spectrum
\bea\label{relation}
k\omega&=&{l}+2(n+1)+|k\lambda-(2s+1)m+{\bar m}|+(2s+1)m+{\bar m}\,.
\eea
We will now see that this spectrum of excitations is reproduced for the corresponding CFT state.

\subsection{The excitation spectrum in the CFT}

\subsubsection{Construction of the CFT state}

\label{SecBldBl}

The scalar excitations discussed in the last section arise from gravitons $h_{ij}$ or gauge fields $C^{(2)}_{ij}$ where the indices $i, j$ lie on the $T^4$. To relate these perturbations with the CFT, we consider the combinations
\be
S^\pm_{ij}=h_{ij}\pm C^{(2)}_{ij} \,.
\ee
For $S^+_{ij}$  the index $i$ is carried by the left movers and the index $j$ by the right movers; for $S^-$ this is reversed. 
We will focus on the lightest excitations, corresponding to $n=0$ in (\ref{relation}).

A scalar with angular momentum $l$ corresponds to the CFT operator which joins together $l+1$ component strings, and which descends from a chiral primary by application of the `anomaly free' part of the conformal algebra. If the initial state had all component strings of winding $k$, then the final state contains a component string of winding
\be
k'=(l+1) k \,.
\label{length}
\ee
An explicit construction of the vertex operator for the CFT excitation is presented in \cite{Avery:2009tu}, and here we just quote the result\footnote{See equation (4.13) in \cite{Avery:2009tu}, which also gives the values of the normalization constant ${\cal C}$.}
\begin{equation}\label{ACM}
\widetilde{\mathcal{V}}^{A\dot{A}B\dot{B}}_{l, l-m_1-\bar{m}_1, m_1-\bar{m}_1}(w,{\bar w})
	= 
	{\cal C}
	(J^+_0)^{m_1}(\bar{J}^+_0)^{\bar{m}_1}
	G^{+A}_{-\frac{1}{2}}\psi^{-\dot{A}}_{-\frac{1}{2}}
	\bar{G}^{\dot{+}B}_{-\frac{1}{2}}\bar{\psi}^{\dot{-}\dot{B}}_{-\frac{1}{2}}
	\tilde{\sigma}^{--}_{l+1}(w,{\bar w}).
\end{equation}
Quantum numbers $m_1$ ${\bar m}_1$ take values between $0$ and $l$, and they are related with 
$m$ and ${\bar m}$ appearing in (\ref{relation}):
\bea
m_1=m+\frac{l}{2},\quad {\bar m}_1={\bar m}+\frac{l}{2}.
\eea
To match the energy of the CFT excitation (\ref{ACM}) with the gravity expression (\ref{relation}), we begin with discussing various features of the initial state (\ref{FrSpFlSt}) and its  perturbation by (\ref{ACM}):
\bea\label{VactOnPsi}
\widetilde{\mathcal{V}}^{A\dot{A}B\dot{B}}_{l, l-m_1-\bar{m}_1, m_1-\bar{m}_1}(w,{\bar w})|\Psi_k\rangle_s
\eea
It is convenient to do a separate analysis of various ingredients of (\ref{ACM}).

\begin{enumerate}[(i)]
\item{
The twist operator $\tilde{\sigma}^{--}_{l+1}$ has $l$ fermions on the left and $l$ fermions on the right, and it adds or removes fermions in a way that leaves the left and the right Fermi seas fully filled. To see this explicitly, we observe that 
$\tilde{\sigma}^{--}_{l+1}$ changes the quantum numbers of the state in the CFT:
\be
\Delta m=-{l\over 2}, ~~\Delta \bar m =-{l\over 2}
\label{hel}
\ee
This change is caused by taking $l+1$ component strings with base spins $m=\bar m=\h$ and joining them into one component string with base spin $m=\bar m =\h$. If we do not remove any fermions that were already present on the component strings, then the quantum numbers have changed as
\bea
\Delta m = \h - (l+1)\h = - {l\over 2}, ~~~\Delta \bar m = \h - (l+1)\h = - {l\over 2}\nonumber
\eea
which agrees with (\ref{hel}). 
}
\item{The twist $\tilde{\sigma}^{--}_{l+1}$ changes the energy of fermions.  The right sector had no fermionic excitations, so the  change in energy from fermions is $\Delta \bar h_f=0$. But in the left sector each of the $l+1$ component strings carried fermions $\psi^+, \t\psi^+$ filling levels ${1\over k}, {2\over k}, \dots {s\over k}$. Thus these fermions had an energy
\bea
h_f^{initial}=(l+1)\times 2\times {1\over k}[1+2+\dots s]={(l+1)\over k} s(s+1)\nonumber \,.
\eea
After the twist, we have $(l+1)s$ fermions of each type on a component string of length (\ref{length}), so we get an energy
\bea\label{DelFermInt}
h^{final}_f=2\times {1\over (l+1) k} [1+2+\dots (l+1) s]={s((l+1)s+1)\over k}\,.
\eea
Thus the change in dimension from these fermions is 
\be\label{DelhFerm}
\Delta h_f = h^{final}_f-h^{initial}_f=-{ls\over k} \,.
\ee
}
\item{The perturbation (\ref{ACM}) excites bosonic fields via
the action of supercurrents $G^{+A}_{-\frac{1}{2}}\bar{G}^{\dot{+}B}_{-\frac{1}{2}}$.  

Since the bosonic field $X$ appears in the supercurrent only through the sum $\sum_i \p X^{(i)}$ over the fields $\p X^{(i)}$ on all the strands of the components string involved in the twisting, we can only generate modes of $\p X$ that are symmetric under the permutation of the $l+1$ strands. Thus the energy of bosonic modes is quantized in the units of ${1\over k}$ rather than  
${1\over {(l+1) k}}$, so we have
\bea\label{SumHb}
\Delta h_b=\frac{r}{k},\quad \Delta {\bar h}_b=\frac{\bar r}{k},\qquad r,{\bar r}\in \mZ \,, \quad r,{\bar r}\ge 1 \,.
\eea
}
\item{ 
Application of $({\bar J}^+_0)^{{\bar m}_1}=[\int {\bar J}^+ d{\bar z}]^{{\bar m}_1}$ to ${\tilde\sigma}^{--}_{l+1}$ contributes the following amount to the  
conformal dimension of (\ref{VactOnPsi}):
\bea\label{hBarJ}
\Delta {\bar h}_J=\frac{{\bar m}_1}{k}.
\eea 
This can be seen as follows. In Appendix~\ref{Sec:AppCFT} it is shown that 
${\bar J}^+_0$ has the effect of raising the dimension by a multiple of $\frac{1}{k}$. The lowest excitation is obtained by ${\bar J}^+_0$ acting as if it were ${\bar J}^+_{-\frac{1}{k}}$. Note that ${\bar J}^+_{-\frac{1}{k}}$ takes the form
\bea
\bar J^+_{-{1\over k}} \sim \bar\psi^+_{-{1\over (l+1)k}}\t{\bar\psi}^+_{-{1\over k }+ {1\over (l+1)k}}+\dots + {\bar\psi}^+_{-{1\over k} + {1\over (l+1)k}}\t {\bar\psi}^+_{-{1\over (l+1)k}}.
\eea
The summation contains $l$ terms, so ${\bar J}^+_{-{1\over k}}$ 
can be applied $l$ times before we encounter constraints from the Pauli exclusion principle, and (\ref{ACM}) has only  
${\bar m}_1\le l$ insertions.}
\item{ 
Application of $(J^+_0)^{m_1}=[\int {J}^+ d{\bar z}]^{{m}_1}$ to ${\tilde\sigma}^{--}_{l+1}$ contributes the following amount to the  
conformal dimension of (\ref{VactOnPsi}):
\bea\label{delHJ}
\Delta {h}_J=\frac{{m}_1(2s+1)}{k}.
\eea 
As in point (iv), $J^+_0$ has the effect of raising the dimension by a multiple of $\frac{1}{k}$.
However, the Fermi sea on the component string of length $(l+1) k$ is already filled to $(l+1) s$ levels for each type of fermion; i.e., it is filled to the levels ${s\over k}$. Thus the lowest energy accessible in a multiple of ${1\over k}$ is ${(2s+1)\over k}$, and this leads to (\ref{delHJ}).
}
\item{
Finally, we have to take into account an important constraint: the total momentum of the CFT state must be an integer. This implies
\be\label{hMinHb}
h-\bar h=\lambda= n_p, ~~~~n_p\in {\mZ}.
\ee
}
\end{enumerate}

\subsubsection{Charges of the CFT state and matching to gravity}

We now combine the ingredients (i)--(vi) to find the energy of the perturbation (\ref{ACM}) and to demonstrate perfect agreement with the gravity expression (\ref{relation}).

We first expand the level-matching condition \eq{hMinHb}, to find
\be\label{DeltaHb-2}
\Delta \bar h_b+\Delta \bar h_J+\la~=~
\Delta h_f+ \Delta h_b+\Delta h_J.
\ee
From point (iii) above, we have
\be
\Delta h_b \ge \frac{1}{k} \,, \qquad \Delta \bar h_b \ge \frac{1}{k} \,.
\ee
We are looking for the quantum numbers of the state of lowest energy. We can find these by either setting $\Delta \bar h_b = \frac{1}{k}$, which then determines $\Delta h_b$ by level-matching, or vice versa.

So we find two different cases: 
\begin{enumerate}
	\item $\Delta h_b \ge \Delta \bar h_b$
	\item $\Delta h_b < \Delta \bar h_b$ .
\end{enumerate}
Equivalently, we can distinguish the cases using the following discriminant:
\bea
D &=& k \left(\Delta h_b - \Delta \bar h_b \right) \cr
  &=& k \left( \Delta \bar h_J + \lambda - \Delta h_f - \Delta h_J \right) \cr
  &=& k\lambda -(2s+1) m + \bar m \,.
\eea
Note that $D$ is precisely the quantity which appears inside the absolute value in the supergravity spectrum~\eq{relation}.
In the second line above we used \eq{DeltaHb-2} and in the third line we used \eq{DelhFerm}, \eq{hBarJ} and \eq{delHJ}.
We next separately consider $D \ge 0$ and $D<0$.

\b

\noindent
{\bf 1. Positive branch
}\\
We first consider $D\ge0$ and combine the ingredients (i)--(vi) to find the lowest available state.
\begin{enumerate}[(a)]
\item{In the right sector, the lightest excitation adds $\bar m_1$ units of $\int {\bar J}^+ d{\bar z}$ and introduces the lightest bosonic mode (${\bar r}=1$ in (\ref{SumHb})). Then combining (\ref{SumHb}) and 
(\ref{hBarJ}) gives
\be\label{DeltaHb-1}
\Delta \bar h ~=~ \Delta \bar h_b+\Delta \bar h_J
~=~{\bar m_1 + 1\over k} \,.
\ee
}
\item{In the left sector, the excitation adds $m_1$ units of $\int {J}^+ d{z}$, changes the energy of the fermions according to (\ref{DelhFerm}), and introduces a bosonic mode with energy $\Delta h_b$. Using (\ref{delHJ}), we 
find
\be\label{DeltaHNb}
\Delta h=\Delta h_f+ \Delta h_b+\Delta h_J=-{sl\over k}+\Delta h_b+{m_1(2s+1)\over k}\,.
\ee
}
\item{Now $\Delta h_b$ is determined uniquely by the level-matching condition
\be
\Delta h=\Delta \bar h+\la \,.
\label{hsevent-1}
\ee
}
\item{We find the CFT energy from \eq{DeltaHb-1} and \eq{hsevent-1}:
\be
\omega_{CFT}~=~2\Delta \bar h+\la ~=~\frac{l+2+2\bar m}{k}+\lambda \,.
\label{hfift}
\ee
To compare to supergravity, we examine (\ref{relation}) with $n=0$ and $D\ge 0$, finding
\be
\omega~=~{l+2+2\bar m\over k}+\la,
\ee
in perfect agreement. 
}
\end{enumerate}

\noindent
{\bf 2. Negative branch }\\
We next consider the opposite case, $D<0$. Now we find the lowest energy state by setting $\Delta h_b={1\over k}$, which determines $\Delta \bar h_b$ by level-matching. In this situation we get
\bea
\Delta h &=& \Delta h_f+\Delta h_b+\Delta h_J~=~ -{ls\over k}+{1\over k} +m_1{(2s+1)\over k} \,, \cr
\Delta{\bar h}&=&\Delta h-\la \,.
\eea
The CFT energy is then
\bea
\omega_{CFT}~=~2\Delta h-\la &=&{l+2 +2(2s+1) m \over k}-\lambda\,. 
\label{hninet}
\eea
To compare to supergravity, we examine (\ref{relation}) with $n=0$ and $D<0$, finding
\bea
\omega&=&{l+2 +2(2s+1) m\over k}-\lambda \,,
\eea
again in exact agreement.

\section{Discussion} \label{sec:discuss}

In recent work we have addressed the question of identifying CFT states which can be thought of as describing `neck' or `cap' degrees of freedom in the dual gravitational description. The cap is the region where the naive black hole horizon appears in the traditional black hole metric.
We found that chiral algebra generators acting on Ramond ground states described gravitational perturbations in the neck region~\cite{\mtone,\mttwo,*\lmt}.

In this paper we have studied CFT states which describe cap degrees of freedom. These states are represented by Fermi seas filled to a level $s/k$, and for generic $s$ cannot be obtained by acting on a Ramond ground state with operators in the chiral algebra. It is natural to conjecture that it may be a general feature that such states describe cap degrees of freedom.

We have identified the gravitational configurations dual to this class of states. These solutions have appeared in the literature,
%
and include both regular solutions as well as solutions with orbifold singularities.
This nontrivial structure in the cap region is controlled by common divisors between the integers $(k,s,s+1)$~\cite{Jejjala:2005yu}.
We have provided strong evidence for this identification by comparing the scalar excitation spectrum between gravity and CFT, and finding exact agreement.

Further support to the above relation between chiral algebra generators and neck degrees of freedom comes from the following observation.
The supergravity solutions with parameters $s$ and $s'=nk+s$ have the same orbifold structure in their cap regions, but differ in the neck region. 
The corresponding CFT states are related by spectral flow by an even integer $\a=2n$, and
this operation can be written terms of the action of chiral algebra generators.

The supergravity configurations contain a spectral flow parameter which can take fractional values $s/k$.
As well as being the CFT Fermi level, there is a more direct interpretation:
the dual CFT states may be constructed from a Ramond vacuum by applying a spectral flow transformation by $\a=2s$ in the covering space of the CFT.

It should be straightforward to extend our results to the case when we include both left and right spectral flow with fractional parameters. The CFT states in this case are the obvious generalizations of those considered in this paper, and it is natural to expect that the dual geometries are non-supersymmetric orbifolded JMaRT solutions~\cite{Jejjala:2005yu}. It would be interesting to explore this further.

The identification proposed in this paper completes the relation between gravity and CFT for the general two-centered Bena-Warner geometry. 
It is a long-standing problem to identify CFT duals of more general Bena-Warner geometries, when such duals exist.
It would be interesting to see if the current work sheds light on possible CFT duals of Bena-Warner geometries with three or more centers, in particular those referred to as `deep' microstates~\cite{Bena:2006kb,*Bena:2007qc}.


\section*{Acknowledgements}

We thank R. Russo for discussions.
The work of SDM and DT  was supported in part by DOE grant DE-FG02-91ER-40690.
The work of SG was partially supported by MIUR-PRIN contract 2009-KHZKRX, by the Padova University Project CPDA119349 and by INFN.



\begin{appendix}

\section{Axially symmetric microstate geometries} \label{app:BW}

In this Appendix we give a concise but self-contained review of the general class of 
 $U(1)\times U(1)$ symmetric microstate geometries constructed in \cite{\bw,Bena:2005va,Berglund:2005vb} and we will examine the most general two-center solution.

The type IIB solutions can always be put in the form
\bea\label{sugra}
ds^2&=& \frac{1}{\sqrt{Z_1 Z_2}}\,\Bigl[-\frac{1}{Z_3} (dt+k)^2 +
Z_3 \Bigl(dy+dt -\frac{dt +k}{Z_3}+a_3\Bigr)^2\Bigr] \nonumber\\
&&+ \sqrt{Z_1 Z_2} \,ds^2_4 +
\sqrt{\frac{Z_1}{Z_2}}\, \sum_{a=1}^4 dz_a^2\,,\nonumber\\
C^{(2)}&=&-\frac{1}{Z_1}\,(dt+k)\wedge (dy+dt +a_3)+ a_1\wedge(dy+dt +a_3)+\gamma_2\,,\nonumber\\
e^{2\Phi}&=&\frac{Z_1}{Z_2}\,.
\eea
 $ds^2_4$ is a metric on the four spatial non-compact directions, $Z_1$, $Z_2$ and $Z_3$ are functions on this 4D space, $k$, $a_1$ and $a_3$ are 1-forms and $\gamma_2$ is a 2-form that is related to the previous objects by\footnote{We denote by $*_4$ the Hodge dual with respect to $ds^2_4$. Similarly $*_3$ will denote the dual with respect to the metric $\mathbb{R}^3$, $ds^2_3$, introduced below. Bold-face letters denote 1-forms on $\mathbb{R}^3$.}
 \be
 d\gamma_2 = *_4 dZ_2 + a_1\wedge da_3\,.
 \ee
 
 Supersymmetry and the $U(1)\times U(1)$ axial isometry imply that 
 $ds^2_4$ is of the Gibbons-Hawking form
 \be
 ds_4^2= V^{-1}\,(d\tau + \mathbf{A})^2 + V\,ds^2_3\,,
 \ee
 with $ds^2_3$ the flat metric on $\mathbb{R}^3$, $V$ a harmonic function on $\mathbb{R}^3$ and $\mathbf{A}$ the 1-form on $\mathbb{R}^3$ dual to $V$: $*_3 d \mathbf{A} = dV$. The coordinate $\tau$ is related to the $\psi$ and $\phi$ Cartan angles used in (\ref{stringmetric}) as $\tau = \phi-\psi$. 
 
The supergravity equations imply that all the other metric functions appearing in (\ref{sugra}) can be explicitly constructed in terms of harmonic functions on $\mathbb{R}^3$. Let us first introduce a third 1-form $a_2$, via the relation
\be
dk+*_4 dk = Z_1\, da_1 + Z_2\, da_2 +Z_3\, da_3\,.
\ee
The three 1-forms $a_I$ ($I=1,2,3$) can be written as
\be
\label{aI}
a_I = \frac{K_I}{V}\,(d\tau + \mathbf{A}) + \mathbf{a}_I\,,\quad *_3 d \mathbf{a}_I = - dK_I\,,
\ee
for some harmonic functions $K_I$ on $\mathbb{R}^3$. Introducing three more harmonic functions $L_I$ allows one to write the three warp factors $Z_I$ as
\be
Z_I = L_I + \frac{|\epsilon_{IJK}|}{2}\,\frac{K_J\,K_K}{V}\,.
\ee
Finally the 1-form $k$ is
\be
k = \mu\,(d\tau + \mathbf{A}) + {\boldsymbol \omega}\,,
\ee
with 
\be
\mu = M + \frac{L_I\,K_I}{2 V} + \frac{K_1 K_2 K_3}{V^2}\,,\quad  *_3 d {\boldsymbol \omega}= V dM - M dV +\frac{1}{2}\,(K_I \,dL_I - L_I\,dK_I)\,,
\ee
where $M$ is another harmonic function. 

The ansatz (\ref{sugra}) satisfies the supergravity equations of motion for generic choices of the harmonic functions introduced above, but the requirement of regularity imposes some constraints. Generically, all the harmonic functions should have poles at the same points in $\mathbb{R}^3$, that we will denote by $\mathbf{x}^{(i)}$. Hence one can write
\bea
V = \sum_i \frac{q^{(i)}}{|\mathbf{x}- \mathbf{x}^{(i)}|}\,,\,\,K_I = \sum_i \frac{d^{(i)}_I}{|\mathbf{x}- \mathbf{x}^{(i)}|}\,,\,\, L_I =1+ \sum_i \frac{Q^{(i)}_I}{|\mathbf{x}- \mathbf{x}^{(i)}|}\,,\,\, M = \sum_i \frac{m^{(i)}}{|\mathbf{x}- \mathbf{x}^{(i)}|} .
\eea
The coefficients $q^{(i)}$ should be integers and they should sum to one, so as to guarantee that $ds^2_4$ is asymptotically $\mathbb{R}^4$:
\be
\label{Vpoles}
q^{(i)}\in \mathbb{Z}\,,\quad \sum_i q^{(i)}=1\,.
\ee
The coefficients $d_I^{(i)}$, that capture the ``dipole charges'' of the geometry, are also expressed in terms of integers $k_I^{(i)}$ as
\be
\label{dI}
d_1^{(i)}=\frac{g\,\alpha'}{2\,R}\,k_1^{(i)}\,,\quad d_2^{(i)} = \frac{g\,\alpha'^3}{2\,V\,R}\,k_2^{(i)}\,,\quad d_3^{(i)} = \frac{R}{2}\,k_3^{(i)}\,.
\ee
 Moreover, 
the cancellation of poles in the functions $Z_I$ and $\mu$ fixes the coefficients $Q_I^{(i)}$ and $m^{(i)}$ in terms of $q^{(i)}$ and $d_I^{(i)}$:
\be\label{polecancellation}
Q_I^{(i)}=-\frac{|\epsilon_{IJK}|}{2}\,\frac{d_J^{(i)}\,d_K^{(i)}}{q^{(i)}}\,,\quad m^{(i)} = \frac{1}{2}\,\frac{d_1^{(i)}\,d_2^{(i)}\,d_3^{(i)}}{(q^{(i)})^2}\,.
\ee
Hence the geometry is entirely specified by the set of integers $q^{(i)}$ and $k_I^{(i)}$, and by the positions of the poles $\mathbf{x}^{(i)}$. The points $\mathbf{x}^{(i)}$ are partially constrained by requiring the absence of Dirac-Misner singularities in ${\boldsymbol \omega}$:
\be
\label{bubble}
\sum_{j\not=i} \Pi_1^{(ij)}\,\Pi_2^{(ij)}\,\Pi_3^{(ij)}\,\frac{q^{(i)}\,q^{(j)}}{|\mathbf{x}^{(i)}-\mathbf{x}^{(j)}|} = - \sum_{I=1}^3 d_I^{(i)}\quad \forall i\,,
\ee 
where
\be
\Pi_I^{(ij)} = \frac{d_I^{(j)}}{q^{(j)}}-\frac{d_I^{(i)}}{q^{(i)}}\,.
\ee

 The ``global''  D1, D5 and P charges of the solution, $Q_I$,\footnote{We identify $Q_1$ with the D1 charge, $Q_2$ with the D5 charge, for which we use the more explicit notation $Q_5$ in the body of the paper, and $Q_3$ with the momentum charge, $Q_p$. Analogously, we identity $n_2\equiv n_5$, $n_3\equiv n_p$.} are given by $Q_I = 4\,\sum_i Q_I^{(i)}$; they are expressed in terms of integers $n_I$ as in (\ref{Q1Q5}, \ref{Qp}). The relations (\ref{polecancellation}) imply
\be
\label{global}
n_I = - \frac{|\epsilon_{IJK}|}{2}\,\sum_i \frac{k_J^{(i)}\,k_K^{(i)}}{q^{(i)}}\,.
\ee
The two angular momenta are given by
\bea
\label{angular}
J_L = \frac{1}{2}\,\sum_i \frac{k_1^{(i)}\,k_2^{(i)}\,k_3^{(i)}}{(q^{(i)})^2}\,,\quad J_R= \frac{4\,V\,R}{g^2\,\alpha'^4}\,|\sum_{I,i} d_I^{(i)}\,\mathbf{x}^{(i)}|\,.
\eea

\subsection{The general two-center solution}
\label{app2centers}
Let us apply the construction described above to the case in which the harmonic functions have poles at two points in $\mathbb{R}^3$, that we can take as the origin and the point $\mathbf{c}$. The constraint (\ref{Vpoles}) implies that
\be
V = -\frac{s}{|\mathbf{x}|} +\frac{s +1}{|\mathbf{x}- \mathbf{c}|}\,,
\ee
where $s$ is an integer. As one can easily see from (\ref{aI}), the shift $K_I \to K_I + c_I \,V$, for arbitrary constant $c_I$'s, acts on the 1-forms $a_I$ as the gauge transformation $a_I \to a_I + c_I\,d\tau$, and hence does not change the geometry. One can exploit this gauge freedom to impose the constraints $\sum_i d_I^{(i)} =0$ for $I=1,2,3$. Thus one can take
\be
K_I = d_I \,\Bigl(\frac{1}{|\mathbf{x}|} -\frac{1}{|\mathbf{x}- \mathbf{c}|}\Bigr)\,,
\ee
where the dipole charges $d_I$ are expressed as in (\ref{dI}) in terms of the integers $k_I$. Eqs. (\ref{bubble}) reduce in this case to one independent constraint that fixes the modulus of $\mathbf{c}$:
\be
\label{cmodulus}
|\mathbf{c}| = \frac{1}{s^2(s+1)^2}\,\frac{d_1 d_2 d_3}{d_1+d_2+d_3} = \frac{Q_p}{4\,s (s+1)}\,\eta\,.
\ee
Hence all 2-center solutions are specified by the choice of four integers, $s$ and $k_I$. These integers have to be chosen in such a way that the three global charges given in (\ref{global}) 
\be \label{eq:BWintegers}
n_1 = \frac{k_2\,k}{s \,(s+1)}\,,\quad n_5 = \frac{k_1\,k}{s \,(s+1)}\,,\quad n_p = \frac{k_1\,k_2}{s\,(s+1)}\,,
\ee
be also integers, where we have renamed $k\equiv k_3$ for later convenience. Moreover, using the relations (\ref{angular}) and (\ref{cmodulus}), we can also compute the angular momenta of the solutions:
\be\label{eq:Japp}
J_L = \Bigl(s+\frac{1}{2}\Bigr)\,\frac{n_1\,n_5}{k}\,,\quad J_R = \frac{1}{2}\,\frac{n_1\,n_5}{k}\,.
\ee
The quantization condition of angular momenta requires that 
\bea \label{eq:ncdef}
n_c &\equiv& \frac{n_1n_5}{k} \quad \in ~\mZ \,.
\eea
We are proposing to identify these solutions with CFT states which have $n_c$ component strings, each of winding $k$.

Let us now work out the restrictions imposed by integrality of the D1, D5 and P charges in terms of $n_1$, $n_5$, $s$ and $k$. Rearranging (\ref{eq:BWintegers}), we have
\bea
k_2 &=& n_1 \frac{s(s+1)}{k} \quad \in ~\mZ \label{eq:cond1} \\
k_1 &=& n_5 \frac{s(s+1)}{k} \quad \in ~\mZ \label{eq:cond2} \\
n_p &=& \frac{n_1n_5}{k} \frac{s(s+1)}{k} \quad \in ~\mZ \,. \label{eq:cond3}
\eea
In the dual CFT, $n_1$ and $n_5$ are usually taken to be relatively prime.
In this case, the first two conditions above imply that 
\bea \label{eq:pc}
\frac{s(s+1)}{k} \quad \in ~\mZ \,
\eea
since otherwise on the RHS of (\ref{eq:cond1}) and (\ref{eq:cond2}) there would be a remaining factor in the denominator which would have to divide both $n_1$ and $n_5$.
Then we see that \eq{eq:pc}, together with (\ref{eq:ncdef}) and (\ref{eq:cond3}), says that the momentum on each component string is integer. This agrees with the usual rules of the orbifold CFT.

In \eq{eq:mom-units-d} of the main text we have proposed that in the orbifold CFT, in general the momentum per component string should be quantized in units of 
\bea
\frac{1}{\gcd(n_1,n_5)} \,.
\eea
This prescription automatically satisfies the conditions \eq{eq:cond1}--\eq{eq:cond2}.\footnote{We note that the conditions \eq{eq:cond1}-\eq{eq:cond2} were conjectured in (\cite{Giusto:2004ip}, Eq.~(6.29)), motivated by duality invariance.} The condition \eq{eq:cond3}, that the total momentum is an integer, must still be imposed independently.

Finally we note that the charge $n_p$ and angular momenta $J_L$ and $J_R$ we have obtained in (\ref{eq:cond3})  and (\ref{eq:Japp}) match the ones in (\ref{Qp}) and (\ref{JLJR}): indeed one can show \cite{Giusto:2004kj} that the two-center solution described here is equivalent to the one in (\ref{stringmetric}). 

To conclude, in this paper we have provided strong evidence that the most general two-center microstate geometry is dual to the fractional spectral flow states (\ref{FrSpFlSt}).

\section{Non-uniqueness of states} \label{app:degenerate}

In this appendix we prove the following result:

\vspace{-3mm}
\begin{quote}
Consider an orbifold CFT state $\ket{\psi_1}$ with all component strings of equal length $k$, and $\psi^+,\tilde\psi^+$ Fermi seas filled to level $s/k$, with the following conditions:
\bea
&&0 < k < n_1 n_5 \label{eq:klim} \\
&&0 < s \le k-1 \\
&&\frac{s(s+1)}{k} ~=~ p_c 
\,. \label{eq:pc1}
\eea
Then this state is degenerate. 
\end{quote}
\vspace{-3mm}

\noindent
We prove this by explicitly constructing another state with the same quantum numbers as $\ket{\psi_1}$.

By the assumption \eq{eq:klim}, the number of component strings is $n_c \ge 2$. 
Let us consider two of the component strings, $A$ and $B$. The other component strings (if any) shall play no role.

The component strings $A$ and $B$ each have winding $k$, base spin $+$, and $\psi^+,\tilde\psi^+$ Fermi seas filled to level $s/k$. The momentum per component string is $p_c$, and so the excitation energy (above the ground state energy) of the strings $A$ and $B$ combined is 
\bea
h_{AB} &=& 2 p_c \,.
\eea
We now construct a different state $\ket{\psi_2}$ with the same quantum numbers as $\ket{\psi_1}$. To define $\ket{\psi_2}$ we shall modify the component strings $A$ and $B$ in the following way:
\vspace{-2mm}
\begin{enumerate}
	\item We keep the total $n_c$ fixed and same base spins, so that we have the same $(\bar j,\bar m)$.
	\item We keep the total number of fermions $\psi^+,\tilde\psi^+$ fixed so that we have the same $(j,m)$.
	\item We keep the right movers in the ground state, so that we have the same $\bar h$.
\end{enumerate}
\vspace{-2mm}
We define new component strings $A'$ and $B'$ to replace $A$ and $B$ as follows. We take $A'$ to have winding $2k-1$ and $B'$ to have winding 1. We take all the fermions from both $A$ and $B$ and put them on the new $A'$, which has lower available excitations.
We treat $\psi^+,\tilde\psi^+$ identically; let us discuss only the $\psi^+$ and add in the $\tilde\psi^+$ at the end.
The lowest available state for the $2s$ fermions of type $\psi^+$ on string $A'$ has energy
\bea
h_{\psi} &=& \frac{1}{2k-1} \left[ 2s(s+1) - s \right] . \label{eq:hpsi}
\eea
We first show $h_\psi \le p_c$.
Using \eq{eq:pc1}, the first term in the square bracket on the RHS of \eq{eq:hpsi} equals $2 k p_c$.
For the second term, we have
\bea
s ~\le~ k-1 \qquad &\Rightarrow& \qquad -s ~\le~ -p_c 
\eea
so we can write 
\bea
h_{\psi} &\le&  \frac{1}{2k-1} \left[ 2 k p_c - p_c \right] ~=~ p_c \,. \label{eq:hpsi-2}
\eea
So since the lowest available state has $h_\psi \le p_c$, we can achieve an energy $h_\psi = p_c$ by using higher energy levels, and adding in the $\tilde\psi^+$ we obtain
\bea
h_{A'B'} &=& 2 p_c \,.
\eea
Thus we have identified a different state with the same quantum numbers as $\ket{\psi_1}$.

\section{Properties of the CFT excitations}
\label{Sec:AppCFT}

In Section \ref{SecBldBl} we evaluated the charges of the CFT excitation (\ref{VactOnPsi}) using its properties (i)--(vi). In this appendix we justify the fractionation of energy in units of $\frac{1}{k}$, which was used in items (iii)--(v). We focus on 
${J}^+$ here; ${\bar J}^+$, $\d X$ and $\bar\d X$ can be analyzed in the same way. 

To demonstrate that the application of $J^+_0$ in (\ref{ACM}) increases the dimension of (\ref{VactOnPsi}) by multiples of $\frac{1}{k}$, we perform the following steps.

\begin{enumerate}[(a)]
\item{To find the amplitude leading to a string with winding $k(l+1)$, we focus on the first $k(l+1)$ strings in (\ref{VactOnPsi}). The first $k$ strings are combined into 
$|\Psi^{(1)}_k\rangle_s$ as in (\ref{FrSpFlSt}), the next group of $k$ strings is combined into $|\Psi^{(2)}_k\rangle_s$ and so on. We introduce the convenient notation
\bea\label{PsiSbosA}
|\Psi_{k,l+1}\rangle_s\equiv
\prod_{\alpha=1}^{l+1}
|\Psi^{(\alpha)}_k\rangle_s \,.
\eea}
\item{Application of $J^+_0$ to ${\tilde{\cal V}}$ in 
(\ref{VactOnPsi}) is accomplished by integrating $J^+(z)dz$ around the point $z=w$. 
Since the spin of $|\Psi_{k,l+1}\rangle_s$ points up, this state has regular OPE with $J^+(z)$, and the integration contour in $J^+_0$ can be deformed to encircle both $z=w$ and $z=0$. After performing this deformation for all $m_1$ copies of $J^+_0$ in (\ref{ACM}), we end up with applying each $J^+_0$ to the combined state 
${\tilde\sigma}_{l+1}^{--}(w,{\bar w})|
\Psi_{k,l+1}\rangle_s$.
}
\item{To produce a combined string of length $k(l+1)$, the copies involved in permutations $|\Psi_k\rangle_s$ and ${\tilde\sigma}_{l+1}$ should be arranged in a 
specific way\footnote{For example, if 
all $l+1$ copies are contained in one of the component strings in $|\Psi_k\rangle_s$, then the perturbation (\ref{ACM}) would break this string and leave the other ones unchanged, and we are not interested in this amplitude.}, and the easiest way to ensure this arrangement is to use the covering space. We refer to \cite{\lmone,\lmtwo} for the details of this construction, here we will only need the basic idea. The covering space for the $k$ copies in 
$|\Psi^{(\alpha)}_k\rangle_s$ is obtained by promoting $k$ functions $\phi^{(a)}(z)$ into sections of one multivalued function $\Phi(z)$ and introducing a Riemann surface with coordinate $t=t(z)$, such that $\Phi(t)$ is single-valued\footnote{One can think of $\phi^{(a)}(z)$ as an arbitrary operator in the CFT, e.g. $X_i^{(a)}$ or $\psi^+_{(a)}$.}. This procedure leads to $l+1$ copies of the $c=6$ CFT, and the twist operator ${\tilde\sigma}_{l+1}$ permutes these copies.
}
\item{Standard arguments imply that the state ${\tilde\sigma}_{l+1}^{--}(w,{\bar w})|\Psi_{k,l+1}\rangle_s$ can be written as a Bogolyubov transformation of the Ramond vacuum \cite{Avery:2010er},
\bea\label{BogolOne}
e^{\sum \gamma_{pq}(w)[
\psi^+_{-p}\psi^-_{-q}+
{\tilde\psi}^+_{-p}{\tilde\psi}^-_{-q}]}
|0_{k(l+1)}\rangle_s\,,
\eea
where $p$ and $q$ are quantized in units of 
$\frac{1}{k(l+1)}$, 
and we will now demonstrate that $p+q$ is quantized in units of $\frac{1}{k}$.
}
\begin{enumerate}[(i)]
\item{Rewriting (\ref{BogolOne}) on the covering space introduced in (c), we find a similar expression involving $t=w^{1/k}$, 
\bea\label{BogolTwo}
e^{\sum \gamma_{p'q'}(t)[
\Psi^+_{-p'}\Psi^-_{-q'}+
{\tilde\Psi}^+_{-p'}{\tilde\Psi}^-_{-q'}]}
|{\tilde 0}_{l+1}\rangle_s\,,
\eea
where now $p'$ and $q'$ are quantized in units of  $\frac{1}{l+1}$.} 
\item{The construction of the fractional modes ${\tilde\Psi}^+_{-p'}$ is given in \cite{\lmone,\lmtwo}, here we quote the result\footnote{See equation (2.8) in \cite{\lmtwo}.}:
\bea\label{OpenFracMode}
{\tilde\Psi}^+_{-p'}=\oint 
\frac{dt}{2\pi i} \left( \, \sum_{\alpha=1}^{l+1}
\Psi_{(\alpha)}^+(t)e^{-2\pi ip'(\alpha-1)} \right) t^{-p'-1/2}
\eea
The modes with $p'=\frac{m}{l+1}$ (where $m$ is an integer) acting
on the twisted R vacuum in (\ref{BogolTwo}) are 
single--valued. To see this, we observe the the twisted vacuum interchanges the copies, so as $t\rightarrow te^{2\pi i}$, the terms in rhs of (\ref{OpenFracMode}) are cyclically permuted, leaving ${\tilde\Psi}^+_{-p'}$ invariant. 
}
\item{The argument of the exponential in (\ref{BogolTwo}) must be dimensionless, this implies that 
\bea
\gamma_{p'q'}(t)=C_{p'q'} t^{p'+q'}.
\eea
The argument must also be single--valued, and since ${\tilde\Psi}^+_{-p'}$, ${\tilde\Psi}^-_{-p'}$ already have this property, we conclude that $p'+q'$ in (\ref{BogolTwo}) should be an integer. This implies that in (\ref{BogolOne}), $p+q$ is quantized in units of $\frac{1}{k}$.}
\end{enumerate}
\item{Operator $J^+_0$ has the form
\bea\label{J0Psi}
J^+_0=\sum \psi^+_{-\frac{m}{k(l+1)}}
{\tilde\psi}^+_{\frac{m}{k(l+1)}}
\eea
and when it is applied to the state (\ref{BogolOne}), the annihilation operators must contract with the exponent. In particular, a term with given $p$ in (\ref{J0Psi}) brings down 
\bea
\gamma_{\frac{m}{k(l+1)},\frac{n}{k(l+1)}}(w)
{\tilde\psi}^+_{-\frac{m}{k(l+1)}}
{\tilde\psi}^-_{-\frac{n}{k(l+1)}}
\eea
from the exponent and replaces 
${\tilde\psi}^+_{-\frac{m}{k(l+1)}}$ by 
${\psi}^+_{-\frac{m}{k(l+1)}}$. This increases the dimension by 
\bea
\Delta h_J=\frac{m}{k(l+1)}+\frac{n}{k(l+1)}\,,
\eea
which is quantized in units of $\frac{1}{k}$ according to (d). 
}
\item{Since the first $s(l+1)$ levels are already filled in the state $|0_{k(l+1)}\rangle_s$ (this can be seen by rewriting  (\ref{FrSpFlSt}) for the combined string), $p$ and $q$ should be larger that $\frac{2s}{k}$, so the dimension must go up by at least $\Delta h_J=\frac{2s+1}{k}$. This proves (\ref{delHJ}) and (\ref{hBarJ}) (the result for the anti--holomorphic sector is obtained by setting $s=0$).
}

\end{enumerate}

\end{appendix}


\begin{adjustwidth}{-3mm}{-3mm} 


\bibliographystyle{utphysM}      


\bibliography{microstates}       

\providecommand{\href}[2]{#2}\begingroup\raggedright\begin{mcitethebibliography}{10}

\bibitem{Mathur:2005zp}
S.~D. Mathur, ``{The fuzzball proposal for black holes: An elementary
  review},'' {\em Fortsch. Phys.} {\bf 53} (2005) 793--827,
\href{http://arXiv.org/abs/hep-th/0502050}{{\tt hep-th/0502050}}$\!\!$
\mciteBstWouldAddEndPuncttrue
\mciteSetBstMidEndSepPunct{\mcitedefaultmidpunct}
{\mcitedefaultendpunct}{\mcitedefaultseppunct}\relax
\EndOfBibitem
\bibitem{Bena:2007kg}
I.~Bena and N.~P. Warner, ``{Black holes, black rings and their microstates},''
  {\em Lect. Notes Phys.} {\bf 755} (2008) 1--92,
\href{http://arXiv.org/abs/hep-th/0701216}{{\tt hep-th/0701216}}$\!\!$
\mciteBstWouldAddEndPuncttrue
\mciteSetBstMidEndSepPunct{\mcitedefaultmidpunct}
{\mcitedefaultendpunct}{\mcitedefaultseppunct}\relax
\EndOfBibitem
\bibitem{Skenderis:2008qn}
K.~Skenderis and M.~Taylor, ``{The fuzzball proposal for black holes},'' {\em
  Phys. Rept.} {\bf 467} (2008) 117--171,
\href{http://arXiv.org/abs/0804.0552}{{\tt 0804.0552}}$\!\!$
\mciteBstWouldAddEndPuncttrue
\mciteSetBstMidEndSepPunct{\mcitedefaultmidpunct}
{\mcitedefaultendpunct}{\mcitedefaultseppunct}\relax
\EndOfBibitem
\bibitem{Mathur:2012zp}
S.~D. Mathur, ``{Black Holes and Beyond},''
\href{http://arXiv.org/abs/1205.0776}{{\tt 1205.0776}}$\!\!$
\mciteBstWouldAddEndPuncttrue
\mciteSetBstMidEndSepPunct{\mcitedefaultmidpunct}
{\mcitedefaultendpunct}{\mcitedefaultseppunct}\relax
\EndOfBibitem
\bibitem{Hawking:1974sw}
S.~W. Hawking, ``{Particle Creation by Black Holes},'' {\em Commun. Math.
  Phys.} {\bf 43} (1975)
199--220$\!\!$
\mciteBstWouldAddEndPuncttrue
\mciteSetBstMidEndSepPunct{\mcitedefaultmidpunct}
{\mcitedefaultendpunct}{\mcitedefaultseppunct}\relax
\EndOfBibitem
\bibitem{Hawking:1976ra}
S.~W. Hawking, ``{Breakdown of Predictability in Gravitational Collapse},''
  {\em Phys. Rev.} {\bf D14} (1976)
2460--2473$\!\!$
\mciteBstWouldAddEndPuncttrue
\mciteSetBstMidEndSepPunct{\mcitedefaultmidpunct}
{\mcitedefaultendpunct}{\mcitedefaultseppunct}\relax
\EndOfBibitem
\bibitem{Mathur:2009hf}
S.~D. Mathur, ``{The information paradox: A pedagogical introduction},'' {\em
  Class. Quant. Grav.} {\bf 26} (2009) 224001,
\href{http://arXiv.org/abs/0909.1038}{{\tt 0909.1038}}$\!\!$
\mciteBstWouldAddEndPuncttrue
\mciteSetBstMidEndSepPunct{\mcitedefaultmidpunct}
{\mcitedefaultendpunct}{\mcitedefaultseppunct}\relax
\EndOfBibitem
\bibitem{Lunin:2001jy}
O.~Lunin and S.~D. Mathur, ``{AdS/CFT duality and the black hole information
  paradox},'' {\em Nucl. Phys.} {\bf B623} (2002) 342--394,
\href{http://arXiv.org/abs/hep-th/0109154}{{\tt hep-th/0109154}}$\!\!$
\mciteBstWouldAddEndPuncttrue
\mciteSetBstMidEndSepPunct{\mcitedefaultmidpunct}
{\mcitedefaultendpunct}{\mcitedefaultseppunct}\relax
\EndOfBibitem
\bibitem{Lunin:2002iz}
O.~Lunin, J.~M. Maldacena, and L.~Maoz, ``{Gravity solutions for the D1-D5
  system with angular momentum},''
\href{http://arXiv.org/abs/hep-th/0212210}{{\tt hep-th/0212210}}$\!\!$
\mciteBstWouldAddEndPuncttrue
\mciteSetBstMidEndSepPunct{\mcitedefaultmidpunct}
{\mcitedefaultendpunct}{\mcitedefaultseppunct}\relax
\EndOfBibitem
\bibitem{Lunin:2004uu}
O.~Lunin, ``{Adding momentum to D1-D5 system},'' {\em JHEP} {\bf 04} (2004)
  054,
\href{http://arXiv.org/abs/hep-th/0404006}{{\tt hep-th/0404006}}$\!\!$
\mciteBstWouldAddEndPuncttrue
\mciteSetBstMidEndSepPunct{\mcitedefaultmidpunct}
{\mcitedefaultendpunct}{\mcitedefaultseppunct}\relax
\EndOfBibitem
\bibitem{Giusto:2004id}
S.~Giusto, S.~D. Mathur, and A.~Saxena, ``{Dual geometries for a set of
  3-charge microstates},'' {\em Nucl. Phys.} {\bf B701} (2004) 357--379,
\href{http://arXiv.org/abs/hep-th/0405017}{{\tt hep-th/0405017}}$\!\!$
\mciteBstWouldAddEndPuncttrue
\mciteSetBstMidEndSepPunct{\mcitedefaultmidpunct}
{\mcitedefaultendpunct}{\mcitedefaultseppunct}\relax
\EndOfBibitem
\bibitem{Giusto:2004ip}
S.~Giusto, S.~D. Mathur, and A.~Saxena, ``{3-charge geometries and their CFT
  duals},'' {\em Nucl. Phys.} {\bf B710} (2005) 425--463,
\href{http://arXiv.org/abs/hep-th/0406103}{{\tt hep-th/0406103}}$\!\!$
\mciteBstWouldAddEndPuncttrue
\mciteSetBstMidEndSepPunct{\mcitedefaultmidpunct}
{\mcitedefaultendpunct}{\mcitedefaultseppunct}\relax
\EndOfBibitem
\bibitem{Bena:2004de}
I.~Bena and N.~P. Warner, ``{One ring to rule them all ... and in the darkness
  bind them?},'' {\em Adv. Theor. Math. Phys.} {\bf 9} (2005) 667--701,
\href{http://arXiv.org/abs/hep-th/0408106}{{\tt hep-th/0408106}}$\!\!$
\mciteBstWouldAddEndPuncttrue
\mciteSetBstMidEndSepPunct{\mcitedefaultmidpunct}
{\mcitedefaultendpunct}{\mcitedefaultseppunct}\relax
\EndOfBibitem
\bibitem{Giusto:2004kj}
S.~Giusto and S.~D. Mathur, ``{Geometry of D1-D5-P bound states},'' {\em Nucl.
  Phys.} {\bf B729} (2005) 203--220,
\href{http://arXiv.org/abs/hep-th/0409067}{{\tt hep-th/0409067}}$\!\!$
\mciteBstWouldAddEndPuncttrue
\mciteSetBstMidEndSepPunct{\mcitedefaultmidpunct}
{\mcitedefaultendpunct}{\mcitedefaultseppunct}\relax
\EndOfBibitem
\bibitem{Jejjala:2005yu}
V.~Jejjala, O.~Madden, S.~F. Ross, and G.~Titchener, ``{Non-supersymmetric
  smooth geometries and D1-D5-P bound states},'' {\em Phys. Rev.} {\bf D71}
  (2005) 124030,
\href{http://arXiv.org/abs/hep-th/0504181}{{\tt hep-th/0504181}}$\!\!$
\mciteBstWouldAddEndPuncttrue
\mciteSetBstMidEndSepPunct{\mcitedefaultmidpunct}
{\mcitedefaultendpunct}{\mcitedefaultseppunct}\relax
\EndOfBibitem
\bibitem{Bena:2005va}
I.~Bena and N.~P. Warner, ``{Bubbling supertubes and foaming black holes},''
  {\em Phys. Rev.} {\bf D74} (2006) 066001,
\href{http://arXiv.org/abs/hep-th/0505166}{{\tt hep-th/0505166}}$\!\!$
\mciteBstWouldAddEndPuncttrue
\mciteSetBstMidEndSepPunct{\mcitedefaultmidpunct}
{\mcitedefaultendpunct}{\mcitedefaultseppunct}\relax
\EndOfBibitem
\bibitem{Berglund:2005vb}
P.~Berglund, E.~G. Gimon, and T.~S. Levi, ``{Supergravity microstates for BPS
  black holes and black rings},'' {\em JHEP} {\bf 0606} (2006) 007,
\href{http://arXiv.org/abs/hep-th/0505167}{{\tt hep-th/0505167}}$\!\!$
\mciteBstWouldAddEndPuncttrue
\mciteSetBstMidEndSepPunct{\mcitedefaultmidpunct}
{\mcitedefaultendpunct}{\mcitedefaultseppunct}\relax
\EndOfBibitem
\bibitem{Saxena:2005uk}
A.~Saxena, G.~Potvin, S.~Giusto, and A.~W. Peet, ``{Smooth geometries with four
  charges in four dimensions},'' {\em JHEP} {\bf 04} (2006) 010,
\href{http://arXiv.org/abs/hep-th/0509214}{{\tt hep-th/0509214}}$\!\!$
\mciteBstWouldAddEndPuncttrue
\mciteSetBstMidEndSepPunct{\mcitedefaultmidpunct}
{\mcitedefaultendpunct}{\mcitedefaultseppunct}\relax
\EndOfBibitem
\bibitem{Balasubramanian:2006gi}
V.~Balasubramanian, E.~G. Gimon, and T.~S. Levi, ``{Four Dimensional Black Hole
  Microstates: From D-branes to Spacetime Foam},'' {\em JHEP} {\bf 0801} (2008)
  056,
\href{http://arXiv.org/abs/hep-th/0606118}{{\tt hep-th/0606118}}$\!\!$
\mciteBstWouldAddEndPuncttrue
\mciteSetBstMidEndSepPunct{\mcitedefaultmidpunct}
{\mcitedefaultendpunct}{\mcitedefaultseppunct}\relax
\EndOfBibitem
\bibitem{Bena:2006kb}
I.~Bena, C.-W. Wang, and N.~P. Warner, ``{Mergers and Typical Black Hole
  Microstates},'' {\em JHEP} {\bf 11} (2006) 042,
\href{http://arXiv.org/abs/hep-th/0608217}{{\tt hep-th/0608217}}$\!\!$
\mciteBstWouldAddEndPuncttrue
\mciteSetBstMidEndSepPunct{\mcitedefaultmidpunct}
{\mcitedefaultendpunct}{\mcitedefaultseppunct}\relax
\EndOfBibitem
\bibitem{Bena:2007qc}
I.~Bena, C.-W. Wang, and N.~P. Warner, ``{Plumbing the Abyss: Black Ring
  Microstates},'' {\em JHEP} {\bf 07} (2008) 019,
\href{http://arXiv.org/abs/0706.3786}{{\tt 0706.3786}}$\!\!$
\mciteBstWouldAddEndPuncttrue
\mciteSetBstMidEndSepPunct{\mcitedefaultmidpunct}
{\mcitedefaultendpunct}{\mcitedefaultseppunct}\relax
\EndOfBibitem
\bibitem{Kanitscheider:2007wq}
I.~Kanitscheider, K.~Skenderis, and M.~Taylor, ``{Fuzzballs with internal
  excitations},'' {\em JHEP} {\bf 06} (2007) 056,
\href{http://arXiv.org/abs/0704.0690}{{\tt 0704.0690}}$\!\!$
\mciteBstWouldAddEndPuncttrue
\mciteSetBstMidEndSepPunct{\mcitedefaultmidpunct}
{\mcitedefaultendpunct}{\mcitedefaultseppunct}\relax
\EndOfBibitem
\bibitem{Ford:2006yb}
J.~Ford, S.~Giusto, and A.~Saxena, ``{A class of BPS time-dependent 3-charge
  microstates from spectral flow},'' {\em Nucl. Phys.} {\bf B790} (2008)
  258--280,
\href{http://arXiv.org/abs/hep-th/0612227}{{\tt hep-th/0612227}}$\!\!$
\mciteBstWouldAddEndPuncttrue
\mciteSetBstMidEndSepPunct{\mcitedefaultmidpunct}
{\mcitedefaultendpunct}{\mcitedefaultseppunct}\relax
\EndOfBibitem
\bibitem{Bena:2010gg}
I.~Bena, N.~Bobev, S.~Giusto, C.~Ruef, and N.~P. Warner, ``{An
  Infinite-Dimensional Family of Black-Hole Microstate Geometries},'' {\em
  JHEP} {\bf 1103} (2011) 022, \href{http://arXiv.org/abs/1006.3497}{{\tt
  1006.3497}}\relax
\mciteBstWouldAddEndPuncttrue
\mciteSetBstMidEndSepPunct{\mcitedefaultmidpunct}
{\mcitedefaultendpunct}{\mcitedefaultseppunct}\relax
\EndOfBibitem
\bibitem{Bena:2011uw}
I.~Bena, J.~de~Boer, M.~Shigemori, and N.~P. Warner, ``{Double, Double
  Supertube Bubble},'' {\em JHEP} {\bf 10} (2011) 116,
\href{http://arXiv.org/abs/1107.2650}{{\tt 1107.2650}}$\!\!$
\mciteBstWouldAddEndPuncttrue
\mciteSetBstMidEndSepPunct{\mcitedefaultmidpunct}
{\mcitedefaultendpunct}{\mcitedefaultseppunct}\relax
\EndOfBibitem
\bibitem{Giusto:2011fy}
S.~Giusto, R.~Russo, and D.~Turton, ``{New D1-D5-P geometries from string
  amplitudes},'' {\em JHEP} {\bf 11} (2011) 062,
\href{http://arXiv.org/abs/1108.6331}{{\tt 1108.6331}}$\!\!$
\mciteBstWouldAddEndPuncttrue
\mciteSetBstMidEndSepPunct{\mcitedefaultmidpunct}
{\mcitedefaultendpunct}{\mcitedefaultseppunct}\relax
\EndOfBibitem
\bibitem{Giusto:2012gt}
S.~Giusto and R.~Russo, ``{Adding new hair to the 3-charge black ring},''
\href{http://arXiv.org/abs/1201.2585}{{\tt 1201.2585}}$\!\!$
\mciteBstWouldAddEndPuncttrue
\mciteSetBstMidEndSepPunct{\mcitedefaultmidpunct}
{\mcitedefaultendpunct}{\mcitedefaultseppunct}\relax
\EndOfBibitem
\bibitem{Dabholkar:2010rm}
A.~Dabholkar, J.~Gomes, S.~Murthy, and A.~Sen, ``{Supersymmetric Index from
  Black Hole Entropy},'' {\em JHEP} {\bf 04} (2011) 034,
\href{http://arXiv.org/abs/1009.3226}{{\tt 1009.3226}}$\!\!$
\mciteBstWouldAddEndPuncttrue
\mciteSetBstMidEndSepPunct{\mcitedefaultmidpunct}
{\mcitedefaultendpunct}{\mcitedefaultseppunct}\relax
\EndOfBibitem
\bibitem{Strominger:1996sh}
A.~Strominger and C.~Vafa, ``{Microscopic Origin of the Bekenstein-Hawking
  Entropy},'' {\em Phys. Lett.} {\bf B379} (1996) 99--104,
\href{http://arXiv.org/abs/hep-th/9601029}{{\tt hep-th/9601029}}$\!\!$
\mciteBstWouldAddEndPuncttrue
\mciteSetBstMidEndSepPunct{\mcitedefaultmidpunct}
{\mcitedefaultendpunct}{\mcitedefaultseppunct}\relax
\EndOfBibitem
\bibitem{Maldacena:1997re}
J.~M. Maldacena, ``{The large N limit of superconformal field theories and
  supergravity},'' {\em Adv. Theor. Math. Phys.} {\bf 2} (1998) 231--252,
\href{http://arXiv.org/abs/hep-th/9711200}{{\tt hep-th/9711200}}$\!\!$
\mciteBstWouldAddEndPuncttrue
\mciteSetBstMidEndSepPunct{\mcitedefaultmidpunct}
{\mcitedefaultendpunct}{\mcitedefaultseppunct}\relax
\EndOfBibitem
\bibitem{Arutyunov:1997gt}
G.~Arutyunov and S.~Frolov, ``{Virasoro amplitude from the S**N R**24 orbifold
  sigma model},'' {\em Theor.Math.Phys.} {\bf 114} (1998) 43--66,
\href{http://arXiv.org/abs/hep-th/9708129}{{\tt hep-th/9708129}}$\!\!$
\mciteBstWouldAddEndPuncttrue
\mciteSetBstMidEndSepPunct{\mcitedefaultmidpunct}
{\mcitedefaultendpunct}{\mcitedefaultseppunct}\relax
\EndOfBibitem
\bibitem{Arutyunov:1997gi}
G.~Arutyunov and S.~Frolov, ``{Four graviton scattering amplitude from S**N
  R**8 supersymmetric orbifold sigma model},'' {\em Nucl.Phys.} {\bf B524}
  (1998) 159--206,
\href{http://arXiv.org/abs/hep-th/9712061}{{\tt hep-th/9712061}}$\!\!$
\mciteBstWouldAddEndPuncttrue
\mciteSetBstMidEndSepPunct{\mcitedefaultmidpunct}
{\mcitedefaultendpunct}{\mcitedefaultseppunct}\relax
\EndOfBibitem
\bibitem{deBoer:1998ip}
J.~de~Boer, ``{Six-dimensional supergravity on S**3 x AdS(3) and 2d conformal
  field theory},'' {\em Nucl. Phys.} {\bf B548} (1999) 139--166,
\href{http://arXiv.org/abs/hep-th/9806104}{{\tt hep-th/9806104}}$\!\!$
\mciteBstWouldAddEndPuncttrue
\mciteSetBstMidEndSepPunct{\mcitedefaultmidpunct}
{\mcitedefaultendpunct}{\mcitedefaultseppunct}\relax
\EndOfBibitem
\bibitem{Dijkgraaf:1998gf}
R.~Dijkgraaf, ``{Instanton strings and hyperKahler geometry},'' {\em
  Nucl.Phys.} {\bf B543} (1999) 545--571,
\href{http://arXiv.org/abs/hep-th/9810210}{{\tt hep-th/9810210}}$\!\!$
\mciteBstWouldAddEndPuncttrue
\mciteSetBstMidEndSepPunct{\mcitedefaultmidpunct}
{\mcitedefaultendpunct}{\mcitedefaultseppunct}\relax
\EndOfBibitem
\bibitem{Seiberg:1999xz}
N.~Seiberg and E.~Witten, ``{The D1/D5 system and singular CFT},'' {\em JHEP}
  {\bf 04} (1999) 017,
\href{http://arXiv.org/abs/hep-th/9903224}{{\tt hep-th/9903224}}$\!\!$
\mciteBstWouldAddEndPuncttrue
\mciteSetBstMidEndSepPunct{\mcitedefaultmidpunct}
{\mcitedefaultendpunct}{\mcitedefaultseppunct}\relax
\EndOfBibitem
\bibitem{Larsen:1999uk}
F.~Larsen and E.~J. Martinec, ``{U(1) charges and moduli in the D1-D5
  system},'' {\em JHEP} {\bf 06} (1999) 019,
\href{http://arXiv.org/abs/hep-th/9905064}{{\tt hep-th/9905064}}$\!\!$
\mciteBstWouldAddEndPuncttrue
\mciteSetBstMidEndSepPunct{\mcitedefaultmidpunct}
{\mcitedefaultendpunct}{\mcitedefaultseppunct}\relax
\EndOfBibitem
\bibitem{David:1999zb}
J.~R. David, G.~Mandal, S.~Vaidya, and S.~R. Wadia, ``{Point mass geometries,
  spectral flow and AdS(3) - CFT(2) correspondence},'' {\em Nucl.Phys.} {\bf
  B564} (2000) 128--141,
\href{http://arXiv.org/abs/hep-th/9906112}{{\tt hep-th/9906112}}$\!\!$
\mciteBstWouldAddEndPuncttrue
\mciteSetBstMidEndSepPunct{\mcitedefaultmidpunct}
{\mcitedefaultendpunct}{\mcitedefaultseppunct}\relax
\EndOfBibitem
\bibitem{Jevicki:1998bm}
A.~Jevicki, M.~Mihailescu, and S.~Ramgoolam, ``{Gravity from CFT on S**N(X):
  Symmetries and interactions},'' {\em Nucl.Phys.} {\bf B577} (2000) 47--72,
\href{http://arXiv.org/abs/hep-th/9907144}{{\tt hep-th/9907144}}$\!\!$
\mciteBstWouldAddEndPuncttrue
\mciteSetBstMidEndSepPunct{\mcitedefaultmidpunct}
{\mcitedefaultendpunct}{\mcitedefaultseppunct}\relax
\EndOfBibitem
\bibitem{Mathur:2011gz}
S.~D. Mathur and D.~Turton, ``{Microstates at the boundary of AdS},'' {\em
  JHEP} {\bf 05} (2012) 014,
\href{http://arXiv.org/abs/1112.6413}{{\tt 1112.6413}}$\!\!$
\mciteBstWouldAddEndPuncttrue
\mciteSetBstMidEndSepPunct{\mcitedefaultmidpunct}
{\mcitedefaultendpunct}{\mcitedefaultseppunct}\relax
\EndOfBibitem
\bibitem{Mathur:2012tj}
S.~D. Mathur and D.~Turton, ``{Momentum-carrying waves on D1-D5 microstate
  geometries},'' {\em Nucl.Phys.} {\bf B862} (2012) 764--780,
\href{http://arXiv.org/abs/1202.6421}{{\tt 1202.6421}}$\!\!$
\mciteBstWouldAddEndPuncttrue
\mciteSetBstMidEndSepPunct{\mcitedefaultmidpunct}
{\mcitedefaultendpunct}{\mcitedefaultseppunct}\relax
\EndOfBibitem
\bibitem{Lunin:2012gp}
O.~Lunin, S.~D. Mathur, and D.~Turton, ``{Adding momentum to supersymmetric
  geometries},''
\href{http://arXiv.org/abs/1208.1770}{{\tt 1208.1770}}$\!\!$
\mciteBstWouldAddEndPuncttrue
\mciteSetBstMidEndSepPunct{\mcitedefaultmidpunct}
{\mcitedefaultendpunct}{\mcitedefaultseppunct}\relax
\EndOfBibitem
\bibitem{Lunin:2001pw}
O.~Lunin and S.~D. Mathur, ``{Three-point functions for M(N)/S(N) orbifolds
  with N = 4 supersymmetry},'' {\em Commun. Math. Phys.} {\bf 227} (2002)
  385--419,
\href{http://arXiv.org/abs/hep-th/0103169}{{\tt hep-th/0103169}}$\!\!$
\mciteBstWouldAddEndPuncttrue
\mciteSetBstMidEndSepPunct{\mcitedefaultmidpunct}
{\mcitedefaultendpunct}{\mcitedefaultseppunct}\relax
\EndOfBibitem
\bibitem{Schwimmer:1986mf}
A.~Schwimmer and N.~Seiberg, ``{Comments on the N=2, N=3, N=4 Superconformal
  Algebras in Two-Dimensions},'' {\em Phys.Lett.} {\bf B184} (1987)
191$\!\!$
\mciteBstWouldAddEndPuncttrue
\mciteSetBstMidEndSepPunct{\mcitedefaultmidpunct}
{\mcitedefaultendpunct}{\mcitedefaultseppunct}\relax
\EndOfBibitem
\bibitem{Brown:1986nw}
J.~D. Brown and M.~Henneaux, ``{Central Charges in the Canonical Realization of
  Asymptotic Symmetries: An Example from Three-Dimensional Gravity},'' {\em
  Commun. Math. Phys.} {\bf 104} (1986)
207--226$\!\!$
\mciteBstWouldAddEndPuncttrue
\mciteSetBstMidEndSepPunct{\mcitedefaultmidpunct}
{\mcitedefaultendpunct}{\mcitedefaultseppunct}\relax
\EndOfBibitem
\bibitem{Lunin:2001ew}
O.~Lunin and S.~D. Mathur, ``{Correlation functions for M(N)/S(N) orbifolds},''
  {\em Int. J. Mod. Phys.} {\bf A16S1C} (2001) 967--969,
\href{http://arXiv.org/abs/hep-th/0006196}{{\tt hep-th/0006196}}$\!\!$
\mciteBstWouldAddEndPuncttrue
\mciteSetBstMidEndSepPunct{\mcitedefaultmidpunct}
{\mcitedefaultendpunct}{\mcitedefaultseppunct}\relax
\EndOfBibitem
\bibitem{Avery:2009xr}
S.~G. Avery and B.~D. Chowdhury, ``{Emission from the D1D5 CFT: Higher
  Twists},'' {\em JHEP} {\bf 1001} (2010) 087,
\href{http://arXiv.org/abs/0907.1663}{{\tt 0907.1663}}$\!\!$
\mciteBstWouldAddEndPuncttrue
\mciteSetBstMidEndSepPunct{\mcitedefaultmidpunct}
{\mcitedefaultendpunct}{\mcitedefaultseppunct}\relax
\EndOfBibitem
\bibitem{Cvetic:1996xz}
M.~Cvetic and D.~Youm, ``{General Rotating Five Dimensional Black Holes of
  Toroidally Compactified Heterotic String},'' {\em Nucl. Phys.} {\bf B476}
  (1996) 118--132,
\href{http://arXiv.org/abs/hep-th/9603100}{{\tt hep-th/9603100}}$\!\!$
\mciteBstWouldAddEndPuncttrue
\mciteSetBstMidEndSepPunct{\mcitedefaultmidpunct}
{\mcitedefaultendpunct}{\mcitedefaultseppunct}\relax
\EndOfBibitem
\bibitem{Cvetic:1997uw}
M.~Cvetic and F.~Larsen, ``{General rotating black holes in string theory: Grey
  body factors and event horizons},'' {\em Phys.Rev.} {\bf D56} (1997)
  4994--5007,
\href{http://arXiv.org/abs/hep-th/9705192}{{\tt hep-th/9705192}}$\!\!$
\mciteBstWouldAddEndPuncttrue
\mciteSetBstMidEndSepPunct{\mcitedefaultmidpunct}
{\mcitedefaultendpunct}{\mcitedefaultseppunct}\relax
\EndOfBibitem
\bibitem{Gutowski:2003rg}
J.~B. Gutowski, D.~Martelli, and H.~S. Reall, ``{All supersymmetric solutions
  of minimal supergravity in six dimensions},'' {\em Class. Quant. Grav.} {\bf
  20} (2003) 5049--5078,
\href{http://arXiv.org/abs/hep-th/0306235}{{\tt hep-th/0306235}}$\!\!$
\mciteBstWouldAddEndPuncttrue
\mciteSetBstMidEndSepPunct{\mcitedefaultmidpunct}
{\mcitedefaultendpunct}{\mcitedefaultseppunct}\relax
\EndOfBibitem
\bibitem{Chowdhury:2008bd}
B.~D. Chowdhury and S.~D. Mathur, ``{Pair creation in non-extremal fuzzball
  geometries},'' {\em Class. Quant. Grav.} {\bf 25} (2008) 225021,
\href{http://arXiv.org/abs/0806.2309}{{\tt 0806.2309}}$\!\!$
\mciteBstWouldAddEndPuncttrue
\mciteSetBstMidEndSepPunct{\mcitedefaultmidpunct}
{\mcitedefaultendpunct}{\mcitedefaultseppunct}\relax
\EndOfBibitem
\bibitem{Lunin:2002fw}
O.~Lunin and S.~D. Mathur, ``{Rotating deformations of AdS(3) x S(3), the
  orbifold CFT and strings in the pp-wave limit},'' {\em Nucl. Phys.} {\bf
  B642} (2002) 91--113,
\href{http://arXiv.org/abs/hep-th/0206107}{{\tt hep-th/0206107}}$\!\!$
\mciteBstWouldAddEndPuncttrue
\mciteSetBstMidEndSepPunct{\mcitedefaultmidpunct}
{\mcitedefaultendpunct}{\mcitedefaultseppunct}\relax
\EndOfBibitem
\bibitem{Avery:2009tu}
S.~G. Avery, B.~D. Chowdhury, and S.~D. Mathur, ``{Emission from the D1D5
  CFT},'' {\em JHEP} {\bf 10} (2009) 065,
\href{http://arXiv.org/abs/0906.2015}{{\tt 0906.2015}}$\!\!$
\mciteBstWouldAddEndPuncttrue
\mciteSetBstMidEndSepPunct{\mcitedefaultmidpunct}
{\mcitedefaultendpunct}{\mcitedefaultseppunct}\relax
\EndOfBibitem
\bibitem{Avery:2010er}
S.~G. Avery, B.~D. Chowdhury, and S.~D. Mathur, ``{Deforming the D1D5 CFT away
  from the orbifold point},'' {\em JHEP} {\bf 06} (2010) 031,
\href{http://arXiv.org/abs/1002.3132}{{\tt 1002.3132}}$\!\!$
\mciteBstWouldAddEndPuncttrue
\mciteSetBstMidEndSepPunct{\mcitedefaultmidpunct}
{\mcitedefaultendpunct}{\mcitedefaultseppunct}\relax
\EndOfBibitem
\end{mcitethebibliography}\endgroup

\end{adjustwidth}

\end{document}